# On the Compression of Translation Operator Tensors in FMM-FFT-Accelerated SIE Simulators via Tensor Decompositions

Cheng Qian and Abdulkadir C. Yucel, *Senior Member, IEEE*

***Abstract*— Tensor decomposition methodologies are proposed to reduce the memory requirement of translation operator tensors arising in the fast multipole method-fast Fourier transform (FMM-FFT)-accelerated surface integral equation (SIE) simulators. These methodologies leverage Tucker, hierarchical Tucker (H-Tucker), and tensor train (TT) decompositions to compress the FFT'ed translation operator tensors stored in three-dimensional (3D) and four-dimensional (4D) array formats. Extensive numerical tests are performed to demonstrate the memory saving achieved by and computational overhead introduced by these methodologies for different simulation parameters. Numerical results show that the H-Tucker-based methodology for 4D array format yields the maximum memory saving while Tucker-based methodology for 3D array format introduces the minimum computational overhead. For many practical scenarios, all methodologies yield a significant reduction in the memory requirement of translation operator tensors while imposing negligible/acceptable computational overhead.**

## I. Introduction

The fast multipole method (FMM)-accelerated surface integral equation (SIE) simulators have become essential tools for the analysis of electromagnetic (EM) scattering from large-scale and complex structures [1, 2]. Among these simulators, the ones exploiting fast Fourier transforms (FFTs) for fast translation stage, also called FMM-FFT-accelerated SIE simulators (FMM-FFT-SIE), have recently received significant attention due to their high parallel scalability and low memory and CPU requirements [3-5]. Compared to the multilevel FMM-accelerated simulators [2], these simulators are easier to implement and parallelize [4] as well as faster for EM analysis of elongated structures [3]. However, just like all SIE simulators, the FMM-FFT-SIE simulators tend to be memory-limited as opposed to CPU-limited. Their memory limitations stem from the memory requirements of the large data structures, including (i) the matrices storing the near-field interactions, (ii) the matrices holding the far-field signatures of basis functions, and (iii) the tensors storing the FFT'ed translation operator samples on a structured grid. To increase the applicability of FMM-FFT-SIE simulators on fixed computational resources, the methodologies for reducing these data structures' memory requirements are called for.

In the past, various compression methodologies were proposed to reduce the memory requirement of these large data structures. In particular, singular value decomposition (SVD) [6], butterfly algorithm [7, 8], and cross algorithm [9] were used in the FMM-accelerated simulators to lessen the memory requirement of near-field interaction matrices while SVD [10] was employed to reduce the memory requirement of matrices storing the far-field signatures. Recently, a tensor compression scheme was proposed to lessen the memory requirement of FFT'ed translation operator tensors [11]. In this scheme, the tensors are compressed via Tucker decompositions [12] during the setup stage of the simulator. Then, the original FFT'ed translation operator tensors are restored from their compressed representations *one-by-one* during the simulators' iterative solution stage [11, 13]. Doing so allows reducing the memory requirement of these tensors more than 90% and introduces negligible computational overhead. Similar Tucker decomposition-based tensor compression strategies have also been used for reducing the memory requirements of systems tensors in FFT-accelerated volume integral equation simulators [14, 15] and FFT-accelerated capacitance [16, 17] and inductance [18] extraction simulators. In addition to Tucker decompositions, tensor train (TT) decompositions [19], were applied to reduce the memory requirements as well as the computational time of EM simulators [20-23]. In such TT-accelerated EM simulators, the main idea is compressing the Toeplitz system matrices via the TT decomposition and performing the matrix-vector multiplications via the reduced-memory TT representations during the iterative solution stage [20-23]. Albeit very promising, these simulators currently necessitate more CPU time compared to FFT-accelerated EM simulators for many realistic problems while still requiring linear (not quasi-linear or logarithmic) memory scaling during the matrix-vector multiplications [21-23].

In [11], Tucker decompositions were applied to each FFT'ed translation operator tensor for each plane-wave direction,

Manuscript received April 28, 2020. This work was supported by Ministry of Education, Singapore, under grant AcRF TIER 1-2018-T1-002-077 (RG 176/18), and the Nanyang Technological University under a Start-Up Grant. (*Corresponding author: Abdulkadir C. Yucel.*)

Cheng Qian and Abdulkadir C. Yucel are with the School of Electrical and Electronic Engineering, Nanyang Technological University, Singapore 639798 (e-mail: cqian@ntu.edu.sg, acyucel@ntu.edu.sg).

stored in a three-dimensional (3D) array format. That said, the performances of Tucker decompositions and hierarchical Tucker (H-Tucker) [24, 25] are not known when applied to all FFT'ed translation operator tensors for all plane-wave directions, stored in four-dimensional (4D) array format. Furthermore, the performance of TT decomposition for reducing the memory requirement of FFT'ed translation operator tensors stored in either 3D or 4D array formats has not been studied yet. Moreover, further research for assessing the performance of the Tucker, H-Tucker, and TT decompositions is needed when the simulation parameters such as the structure size, decomposition tolerance, FMM box size, FMM accuracy, and constitutive parameters are changed.

This paper aims to fill the abovementioned gaps in the application of tensor decompositions to the compression of FFT'ed translation operator tensors. In particular, this paper's contribution is two-fold. First, it introduces the Tucker, H-Tucker, and TT based tensor compression schemes for reducing the memory requirement of FFT'ed translation operator tensors stored in 4D array format. Furthermore, it provides the algorithms for the rapid restoration of the original tensors (3D arrays) from Tucker, H-Tucker, and TT compressed tensors stored in 4D array format. Second, it demonstrates the performance of Tucker, H-Tucker, and TT based methodologies presented for 3D and 4D array formats while the simulation parameters are varied. These simulation parameters include the structure size, decomposition tolerance, FMM box size, FMM accuracy, and constitutive parameters.

The proposed Tucker and TT-based methodologies for compressing 3D and 4D arrays, henceforth called Tucker-3D, TT-3D, Tucker-4D, H-Tucker, and TT-4D methodologies, respectively, have been deployed in an FMM-FFT-SIE simulator. (Note: H-Tucker has only been used to compress 4D arrays as it is an effective generalization of the Tucker for arrays with dimensionality larger than three [24, 25].) The tensor decomposition enhanced simulator has been applied to the analysis of EM scattering from a sphere with $32\lambda$ diameter and a plate with dimensions of $30\lambda \times 42\lambda$, where $\lambda$ is the wavelength. In both practical scenarios, 4D methodologies achieve more than 90% memory reduction while the 3D methodologies yield around 80% memory reduction, for the decomposition tolerance of $10^{-6}$. In the numerical tests, H-Tucker yields the maximum compression, while Tucker-3D requires the minimum computational overhead. With increasing structure size, the memory requirement of the tensors compressed via H-Tucker scales with $O(K^{0.55})$ (quasi-linearly) while the computational overhead introduced by Tucker-3D scales with $O(K \log K)$, where $K$ represents the total number of boxes used to discretize the computational domain. Overall, while the memory saving achieved by the methodologies increases, the computational overhead decreases. The memory saving achieved by all decompositions increases with decreasing FMM box size and increasing structure size, FMM accuracy, decomposition tolerance, and the loss in the medium.

## II. Formulation

In this section, the electric field SIE, its discretization, and its accelerated solution via the FMM-FFT methodology is explained first. (Note: For the sake of brevity in explanation, here only electric field SIE is considered; the reader is referred to [11, 13] for the combined field SIE and other SIEs and their expedient solution via FMM-FFT.) Then, the tensor decompositions, including the proposed Tucker-3D, TT-3D, Tucker-4D, H-Tucker, and TT-4D methodologies, are expounded; the SVD-based algorithms for obtaining Tucker, H-Tucker, and TT decompositions are provided, while the references to the cross-based algorithms [26-29] for efficiently obtaining Tucker and TT decompositions of tensors are referred to. Finally, the algorithms for rapid restoration/decompression are explained to keep the computational overhead minimal during the iterative solution stage of the simulator.

### A. Electric Field SIE and Its FMM-FFT-Accelerated Solution

Let $S$ denote an arbitrarily shaped perfect electric conductor (PEC) surface in a medium with permittivity $\varepsilon$, conductivity $\sigma$, and permeability $\mu$. $S$ is excited by an incident plane-wave with electric field $\mathbf{E}^i(\mathbf{r})$. This electric field induces surface current $\mathbf{J}$ on $S$, which generates scattered electric field $\mathbf{E}^s(\mathbf{r},\mathbf{J})$. The relation between incident and scattered fields on $S$ is formulated via the following electric field SIE

$$\begin{aligned}\hat{\mathbf{n}}\times\hat{\mathbf{n}}\times\mathbf{E}^i(\mathbf{r}) &= -\hat{\mathbf{n}}\times\hat{\mathbf{n}}\times\mathbf{E}^s(\mathbf{r},\mathbf{J}) \\ &= \hat{\mathbf{n}}\times\hat{\mathbf{n}}\times\frac{j\omega\mu}{4\pi}\int_S\left(1+\frac{\nabla\nabla'}{k^2}\cdot\right)G(\mathbf{r},\mathbf{r}')\mathbf{J}(\mathbf{r}')d\mathbf{r}'.\end{aligned} \quad (1)$$

Here $G(\mathbf{r},\mathbf{r}') = \exp(-jk|\mathbf{r}-\mathbf{r}'|)/|\mathbf{r}-\mathbf{r}'|$, $\hat{\mathbf{n}}$ is the outward pointing unit normal to $S$, $k=\omega(\mu\varepsilon)^{0.5}$, $\omega = 2\pi f$, $f$ is the frequency, $\mathbf{r}$ and $\mathbf{r}'$ denote the observation and source locations, respectively. To solve (1), $\mathbf{J}$ is discretized via $N$ Rao-Wilton-Glisson basis functions $\mathbf{b}_n(\mathbf{r})$ [30] as $\mathbf{J}(\mathbf{r}) = \sum_{n=1}^{N} I_n \mathbf{b}_n(\mathbf{r})$. After substituting the discrete representation into (1) and applying Galerkin testing with $\mathbf{b}_m(\mathbf{r})$, $m=1,\ldots,N$, a linear system of equations (LSE) is obtained as

$$\mathbf{V} = \bar{\mathbf{Z}}\cdot\mathbf{I}\ , \quad (2)$$

where $\mathbf{V}_m = \left\langle \mathbf{b}_m, \hat{\mathbf{n}}\times\hat{\mathbf{n}}\times\mathbf{E}^i(\mathbf{r})\right\rangle$, $\bar{\mathbf{Z}}_{m,n} = \left\langle \mathbf{b}_m, -\hat{\mathbf{n}}\times\hat{\mathbf{n}}\times\mathbf{E}^s(\mathbf{r},\mathbf{b}_n)\right\rangle$, $\mathbf{I}_n = I_n$, and $\langle\cdot,\cdot\rangle$ is the standard inner product. Iterative solution of the LSE in (2) requires $O(N^2)$ CPU and memory resources which can be reduced to $O(N^{4/3}\log^{2/3} N)$ via the FMM-FFT scheme [4]. (Note: a transpose-free quasi minimal residual iterative solver [31] with tolerance $10^{-6}$ is used to iteratively solve the LSE in (2).)

In this scheme, the computational domain (a fictitious box) enclosing $S$ is divided into $K_x$, $K_y$, and $K_z$ small boxes along *x*-, *y*-, and *z*- directions; totally $K = K_x K_y K_z$ small boxes are generated. The centers of these boxes labeled by $B_{\mathrm{v}}$ coincide with uniform grid points $\mathbf{r}_{\mathrm{v}}$, where $\mathbf{v}=(v_x,v_y,v_z)$, $v_x=1,\ldots,K_x, v_y=1,\ldots,K_y$, and $v_z=1,\ldots,K_z$. In this scheme, the interactions between basis functions in two nearby non-empty boxes are computed classically if the distance between these boxes, $R_{\mathrm{v'v}}=|\mathbf{R}_{\mathrm{v'v}}|=|\mathbf{r}_{\mathrm{v'}}-\mathbf{r}_{\mathrm{v}}|$, is smaller than $\kappa R^s$, where $R^s$ is the radius of the sphere enclosing the box and $\kappa=4$ in this study. These near-field interactions are stored classically and give rise to the first large data structure referred to in the introduction. On the other hand, the interactions between far non-empty boxes satisfying the condition $R_{\mathrm{v'v}}>\kappa R^s$ are computed as detailed below. First, the basis functions' far-field patterns along a plane-wave direction $\hat{\mathbf{k}}_p$, $p=1,\ldots,N_{\mathrm{dir}}$, which is

$$\mathbf{P}^{\pm}(\hat{\mathbf{k}}_p,\mathbf{b}_n)=\int_{S^b}(\bar{\mathbf{I}}-\hat{\mathbf{k}}_p\hat{\mathbf{k}}_p)\mathbf{b}_n(\mathbf{r})\exp(\pm jk\hat{\mathbf{k}}_p\cdot(\mathbf{r}-\mathbf{r}_{\mathrm{v}}))d\mathbf{r}\,, \quad (3)$$

are summed for each non-empty box as

$$\mathcal{A}_{\mathrm{v}}(\hat{\mathbf{k}}_p)=\sum_{n\in B_{\mathrm{v}}}\mathbf{P}^{+}(\hat{\mathbf{k}}_p,\mathbf{b}_n)I_n. \quad (4)$$

Note that $\mathcal{A}_{\mathrm{v}}(\hat{\mathbf{k}}_p)=0$ for an empty box. In (3), $S^b$ is the support of basis function $\mathbf{b}(\mathbf{r})$, $\hat{\mathbf{k}}_p$, $p=1,\ldots,N_{\mathrm{dir}}$, is pointing along the grid points on a unit sphere, obtained by Cartesian product of quadrature points (see [32, 33] for details). Here, $N_{\mathrm{dir}}=(L+1)(2L+1)$ is the number of plane-wave directions and $L$ is the number of multipoles selected using $L=2kR^s+(2kR^s)^{1/3}1.8(\log_{10}(\Gamma^{-1}))^{2/3}$ [33]; $\Gamma$ is the number of desired accurate digits in the FMM approximation. As $\bar{\mathbf{I}}-\hat{\mathbf{k}}_p\hat{\mathbf{k}}_p=\hat{\boldsymbol{\theta}}\hat{\boldsymbol{\theta}}+\hat{\boldsymbol{\phi}}\hat{\boldsymbol{\phi}}$, only $\theta$ and $\phi$ components of far-field patterns are accounted for. The far-field patterns of all basis functions, $\mathbf{P}^{+}(\hat{\mathbf{k}}_p,\mathbf{b}_n)$, $n=1,\ldots,N$, $p=1,\ldots,N_{\mathrm{dir}}$, are computed and stored during the setup stage of the simulator and constitute the second large data structure mentioned in the introduction. Next, the far-field patterns of all boxes for each plane-wave direction, $\mathcal{A}_{\mathrm{v}}(\hat{\mathbf{k}}_p)$, are convolved with the FFT'ed translation operator tensor $\mathcal{T}_{\mathrm{v'-v}}(\hat{\mathbf{k}}_p)$ using FFT and inverse FFT operators, $\Im(\cdot)$, and $\Im^{-1}(\cdot)$, respectively, as

$$\mathcal{B}_{\mathrm{v'}}(\hat{\mathbf{k}}_p)=\Im^{-1}(\mathcal{T}_{\mathrm{v'-v}}(\hat{\mathbf{k}}_p)\Im(\mathcal{A}_{\mathrm{v}}(\hat{\mathbf{k}}_p))). \quad (5)$$

The plane-wave spectra $\mathcal{B}_{\mathrm{v'}}(\hat{\mathbf{k}}_p)$ resulting from the convolution operation are projected onto each basis function $\mathbf{b}_m$ in $B_{\mathrm{v'}}$ and the contributions to the matrix-vector product are computed via summing over all plane-wave directions as

$$\sum_{p=1}^{N_{dir}}w_p\mathbf{P}^{-}(\hat{\mathbf{k}}_p,\mathbf{b}_m)\mathcal{B}_{\mathrm{v'}}(\hat{\mathbf{k}}_p)\,, \quad (6)$$

where $w_p$, $p=1,\ldots,N_{\mathrm{dir}}$, are the quadrature weights and $\mathbf{P}^{-}(\hat{\mathbf{k}}_p,\mathbf{b}_m)$ is directly obtained by conjugating $\mathbf{P}^{+}(\hat{\mathbf{k}}_p,\mathbf{b}_m)$. The FFT'ed translation operator for each plane-wave direction is obtained via $\mathcal{T}_{\mathrm{v'-v}}(\hat{\mathbf{k}}_p)=\Im(\mathcal{T}^{\sim}_{\mathrm{v'-v}}(\hat{\mathbf{k}}_p))$, where

$$\mathcal{T}^{\sim}_{\mathrm{v'-v}}(\hat{\mathbf{k}}_p)=\frac{-k^2\eta}{16\pi^2}\sum_{l=1}^{L}(-j)^l(2l+1)\Phi_l(\hat{\mathbf{R}}_{\mathrm{v'v}}\cdot\hat{\mathbf{k}}_p)h_l^{(2)}(kR_{\mathrm{v'v}})\,. \quad (7)$$

In the above, $\eta=\sqrt{\mu/\varepsilon}$, $\hat{\mathbf{R}}_{\mathrm{v'v}}=\mathbf{R}_{\mathrm{v'v}}/R_{\mathrm{v'v}}$, $\Phi_l(\cdot)$ denotes the $l^{\mathrm{th}}$-degree Legendre polynomial, and $h_l^{(2)}(\cdot)$ is the $l^{\mathrm{th}}$-order spherical Hankel function of the second kind. The FFT'ed translation operators for all plane-wave directions, $\mathcal{T}_{\mathrm{v'-v}}(\hat{\mathbf{k}}_p)$, $p=1,\ldots,N_{\mathrm{dir}}$, constitute the third largest data structured mentioned in the introduction.

### B. Tensor Decompositions

The FFT'ed translation operators stored in 3D and 4D array formats are compressed via Tucker, H-Tucker, and TT decompositions during the simulator's setup stage. During the iterative solution stage of the simulator, the compressed tensors are (partially or fully) restored to obtain $\mathcal{T}_{\mathrm{v'-v}}(\hat{\mathbf{k}}_p)$ for each direction and perform the convolution in (5). Let $\mathcal{T}_{\mathrm{3D}}$, represent one FFT'ed translation operator tensor for a plane-wave direction, which corresponds to $\mathcal{T}_{\mathrm{v'-v}}(\hat{\mathbf{k}}_p)$, (i.e., a 3D array with dimensions $2K_x\times 2K_y\times 2K_z=n_1\times n_2\times n_3$); there exist $N_{\mathrm{dir}}$ number of $\mathcal{T}_{\mathrm{3D}}$ arrays. Furthermore, let $\mathcal{T}_{4D}$ denote FFT'ed translation operators for all plane-wave directions (i.e., a 4D array with dimensions $2K_x\times 2K_y\times 2K_z\times N_{\mathrm{dir}}=n_1\times n_2\times n_3\times n_4$). These tensors are Tucker, H-Tucker, and TT compressible due to their low rank; their compressed representations can be obtained via the SVD, as detailed below. The original tensors can then be restored from the compressed representations via the methods explained next.

#### 1) Tucker and H-Tucker Decompositions

The tensors $\mathcal{T}_{\mathrm{3D}}$ and $\mathcal{T}_{4D}$ are represented via Tucker decomposition as

$$\mathcal{T}_{\mathrm{3D}}=\mathcal{C}_{\mathrm{T}}\times_1\bar{\mathbf{U}}_{\mathrm{T}}^1\times_2\bar{\mathbf{U}}_{\mathrm{T}}^2\times_3\bar{\mathbf{U}}_{\mathrm{T}}^3, \quad (8)$$

$$\mathcal{T}_{\mathrm{4D}}=\mathcal{C}_{\mathrm{T}}\times_1\bar{\mathbf{U}}_{\mathrm{T}}^1\times_2\bar{\mathbf{U}}_{\mathrm{T}}^2\times_3\bar{\mathbf{U}}_{\mathrm{T}}^3\times_4\bar{\mathbf{U}}_{\mathrm{T}}^4, \quad (9)$$

where $\mathcal{C}_{\mathrm{T}}$ denote the core tensor with dimensions $r_1\times r_2\times r_3$ and $r_1\times r_2\times r_3\times r_4$ for 3D and 4D arrays, respectively. $\bar{\mathbf{U}}_{\mathrm{T}}^i$, $i=1,\ldots,d$, $d=\{3,4\}$ represents the factor matrices with dimensions $n_i\times r_i$. The conceptual representations of these tensors and their Tucker representations are given in Figs. 1(a) and (b). The $\times_i$ represents the $i-$ mode matrix product of a tensor, which is performed for the 1st dimension in (8) as an example as

$$\mathcal{Y}_{v_1,v_2,v_3}=\mathcal{C}_{\mathrm{T}}\times_1\bar{\mathbf{U}}_{\mathrm{T}}^1=\sum_{i'=1}^{r_1}\mathcal{C}_{i',v_2,v_3}\left(\bar{\mathbf{U}}_{\mathrm{T}}^1\right)_{v_1,i'} \quad (10)$$

where $\mathcal{Y}_{v_1,v_2,v_3}$ is the resulting tensor with indices $v_1=1,\ldots,n_1$, $v_2=1,\ldots,r_2$, and $v_3=1,\ldots,r_3$. (Note: The entries of $\bar{\mathbf{U}}_{\mathrm{T}}^1$ are indicated in the subscript next to the parenthesis.) The procedure to obtain the $\mathcal{C}_{\mathrm{T}}$ and $\bar{\mathbf{U}}_{\mathrm{T}}^i$ by utilizing the high-order singular value decomposition (HOSVD) [12] for given tolerance, $\gamma$, is summarized in Algorithm 1 of Appendix. This algorithm requires computationally costly SVDs of mode $-i$ unfolding matrices $\bar{\mathbf{T}}^i$, $i=1,\ldots,d$, which becomes prohibitive for large tensors in large scale problems. To avoid this problem, a 3D Tucker-cross can be considered [28].

As a generalization of the Tucker decomposition for the arrays with high dimensionality ( $d>3$ ), H-Tucker decomposition is used to represent $\mathcal{T}_{4\mathrm{D}}$ as

$$\mathcal{T}_{4\mathrm{D}} = [(\mathcal{C}_{\mathrm{HT}}^{34} \times_3 \bar{\mathbf{C}}_{\mathrm{HT}}^{1234}) \times_3^3 \mathcal{C}_{\mathrm{HT}}^{12}] \times_1 \bar{\mathbf{U}}_{\mathrm{HT}}^1 \times_2 \bar{\mathbf{U}}_{\mathrm{HT}}^2 \times_3 \bar{\mathbf{U}}_{\mathrm{HT}}^3 \times_4 \bar{\mathbf{U}}_{\mathrm{HT}}^4. \quad (11)$$

Here $\bar{\mathbf{C}}_{\mathrm{HT}}^{1234}$ is transfer matrix with dimensions $r_{12}\times r_{34}$. $\mathcal{C}_{\mathrm{HT}}^{12}$ and $\mathcal{C}_{\mathrm{HT}}^{34}$ denote the transfer tensors with dimensions $r_1\times r_2\times r_{12}$ and $r_3\times r_4\times r_{34}$, respectively. $\bar{\mathbf{U}}_{\mathrm{HT}}^i$, $i=1,\ldots,4$, represents the factor matrices with dimensions $n_i\times r_i$ (See the conceptual representation of these tensors/matrices in [Fig. 1(c)]). The symbol $\times_3^3$ in (11) represents tensor contraction along mode-3 of both tensors, which is performed for example as

$$\mathcal{Z}_{v_1,v_2,v_3,v_4} = \mathcal{X} \times_3^3 \mathcal{Y} = \sum_{i'=1}^{r_{34}} \mathcal{X}_{v_1,v_2,i'} \mathcal{Y}_{v_3,v_4,i'}, \quad (12)$$

where $\mathcal{X}$ and $\mathcal{Y}$ are tensors with dimensions $r_1\times r_2\times r_{12}$ and $r_3\times r_4\times r_{12}$, respectively. $\mathcal{Z}_{v_1,v_2,v_3,v_4}$ is the resulting tensor with indices $v_1=1,\ldots,r_1$, $v_2=1,\ldots,r_2$, $v_3=1,\ldots,r_3$, and $v_4=1,\ldots,r_4$. The H-Tucker decomposition can be obtained via an SVD-based methodology provided in Algorithm 2 of Appendix [24, 25].

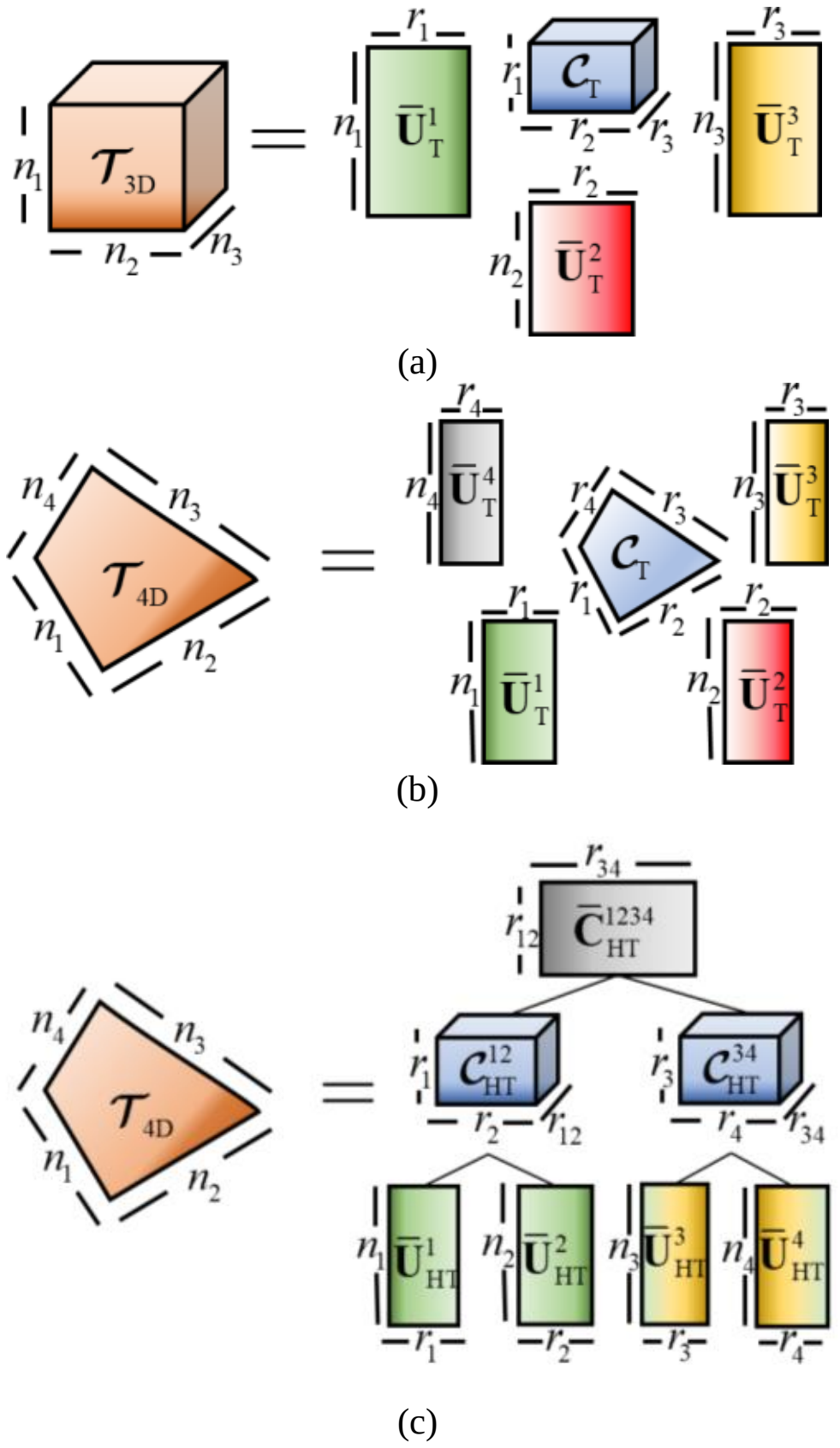

Fig. 1. The conceptual representation of the Tucker decompositions of (a) $\mathcal{T}_{3\mathrm{D}}$ and (b) $\mathcal{T}_{4\mathrm{D}}$ and (c) the H-Tucker decomposition of $\mathcal{T}_{4\mathrm{D}}$. (Note: 4D array is represented by a trapezoidal; each edge of trapezoidal corresponds to one dimension of 4D array. 2D and 3D arrays are represented by a rectangle and cube, respectively. The number of elements along each dimension is specified next to the pertinent edge.)

### 2) *TT Decomposition*

The tensors $\mathcal{T}_{3\mathrm{D}}$ and $\mathcal{T}_{4\mathrm{D}}$ are represented via TT decomposition as

$$\mathcal{T}_{3\mathrm{D}} = \mathcal{C}_{\mathrm{TT}}^1 \times_1 \bar{\mathbf{U}}_{\mathrm{TT}}^1 \times_3 \bar{\mathbf{U}}_{\mathrm{TT}}^3, \quad (13)$$

$$\mathcal{T}_{4\mathrm{D}} = (\mathcal{C}_{\mathrm{TT}}^1 \times_1 \bar{\mathbf{U}}_{\mathrm{TT}}^1) \times_3^1 (\mathcal{C}_{\mathrm{TT}}^2 \times_3 \bar{\mathbf{U}}_{\mathrm{TT}}^3), \quad (14)$$

where $\mathcal{C}_{\mathrm{TT}}^i$, $i=1,\ldots,d-2$, $d=\{3,4\}$ denotes the TT core tensor with dimensions $r_i\times n_{i+1}\times r_{i+1}$ and $\bar{\mathbf{U}}_{\mathrm{TT}}^{1,3}$ represents the factor matrices with dimensions $n_1\times r_1$ and $n_d\times r_{d-1}$, respectively. The conceptual representations of these tensors are provided in Figs. 2(a) and (b). The procedure to obtain the $\mathcal{C}_{\mathrm{TT}}^i$, $i=1,\ldots,d-2$, and $\bar{\mathbf{U}}_{\mathrm{TT}}^{1,3}$ for given tolerance $\gamma$, is provided in Algorithm 3 of Appendix.

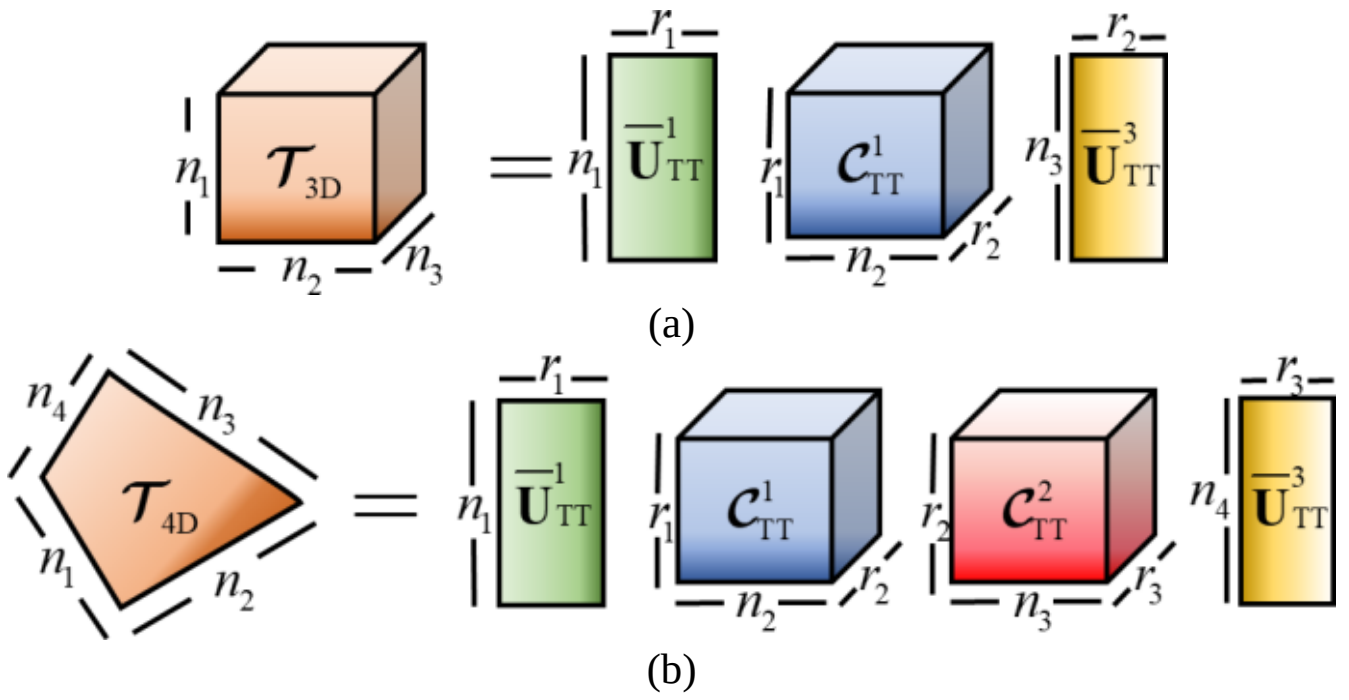

Fig. 2. The conceptual representation of the TT decompositions of (a) $\mathcal{T}_{3\mathrm{D}}$ and (b) $\mathcal{T}_{4\mathrm{D}}$, respectively.

The SVD-based algorithm requires excessive computational resources for obtaining TT decompositions of large tensors. To this end, similar to the Tucker-cross, an adaptive interpolation algorithm called TT-cross provided in [26] can be used. For the numerical results presented below, SVD-based algorithms are efficient enough; the CPU time required to obtain the tensor decomposition is always negligible compared to the CPU time required for the setup and iterative solution stages of the simulator. On the other hand, since tensor restorations are required in each iteration during the iterative solution, efficient tensor restoration schemes, discussed in the next section, are crucial to have minimal computational overhead.

## C. *Tensor Restoration*

Here, tensor restoration schemes are explained for rapid restoration of the original FFT'ed translation operator tensors for each plane-wave direction, $\mathcal{T}_{\mathbf{v}'-\mathbf{v}}(\hat{\mathbf{k}}_p)$, $p=1,\ldots,N_{\mathrm{dir}}$. The original tensors are obtained from the compressed tensors *one-by-one* with minimal computational requirements using these schemes during the iterative solution stage of the FMM-FFT-SIE simulator. Note that such restoration operations, as well as the compression operations, are performed locally in each processor while executing the parallel FMM-FFT-SIE simulators [11], [13]. These operations require no communication among the processors and are performed only for certain plane-wave directions, which the processor is responsible for (see [11] and [13] for details).

*1) Tucker and H-Tucker Restorations*

The original FFT'ed translation operator tensor pertinent to $p^{\text{th}}$ plane-wave direction, $\mathcal{T}_{\mathbf{v}'-\mathbf{v}}(\hat{\mathbf{k}}_p) = \mathcal{T}_{3\text{D}}$, can be obtained from Tucker-compressed tensors via consecutive $i-$ mode matrix products and reshaping operations. These operations are demonstrated in Fig. 3(a) for restoring the original tensors from the core tensor and factor matrices of Tucker-3D methodology. For the Tucker-4D and H-Tucker methodology, the same procedure is applied with an additional step explained in Fig. 3(b) and Fig. 3(c), respectively. For the restoration of Tucker-3D [Fig. 3(a)], the core tensor $\mathcal{C}_{\text{T}}$ is converted to a matrix $\bar{\mathbf{C}}$, which is multiplied by $\bar{\mathbf{U}}_{\text{T}}^{1}$ in Step 1. The resulting matrix $\bar{\mathbf{T}}$ is reshaped and multiplied by $\bar{\mathbf{U}}_{\text{T}}^{2}$ and $\bar{\mathbf{U}}_{\text{T}}^{3}$ in Steps 2 and 3, respectively. In the final step, the resulting matrix $\bar{\mathbf{T}}$ is converted to 3D array $\mathcal{T}_{3\text{D}}$. For the restoration of Tucker-4D and H-Tucker, the core tensor $\mathcal{C}_{\text{T}}$ in the Step 1 of Tucker-3D restoration is obtained by additional steps, respectively. Tucker-4D restoration requires to convert the core tensor $\mathcal{C}_{\text{T}}$ of the Tucker-4D methodology to the matrix $\bar{\mathbf{C}}$ [Fig. 3(b)]. This matrix is then multiplied by the vector obtained by the transpose of $p^{\text{th}}$ row of $\bar{\mathbf{U}}_{\text{T}}^{4}$. The resulting vector is converted to $\mathcal{C}_{\text{T}}$ to be used in Tucker-3D restoration. Similarly, H-Tucker restoration requires the conversion of $\mathcal{C}_{\text{HT}}^{12}$ and $\mathcal{C}_{\text{HT}}^{34}$ to the temporary matrices $\bar{\mathbf{C}}_2$ and $\bar{\mathbf{C}}_1$, respectively [Fig. 3(c)]. After multiplying $\bar{\mathbf{C}}_1$ with the vector obtained by the transpose of $p^{\text{th}}$ row of $\bar{\mathbf{U}}_{\text{T}}^{4}$, the resulting matrix is multiplied with $\bar{\mathbf{C}}_{\text{HT}}^{1234}$ and $\bar{\mathbf{C}}_2$, respectively. The matrix holding the multiplication results is converted to $\mathcal{C}_{\text{T}}$ to be used in Tucker-3D restoration.

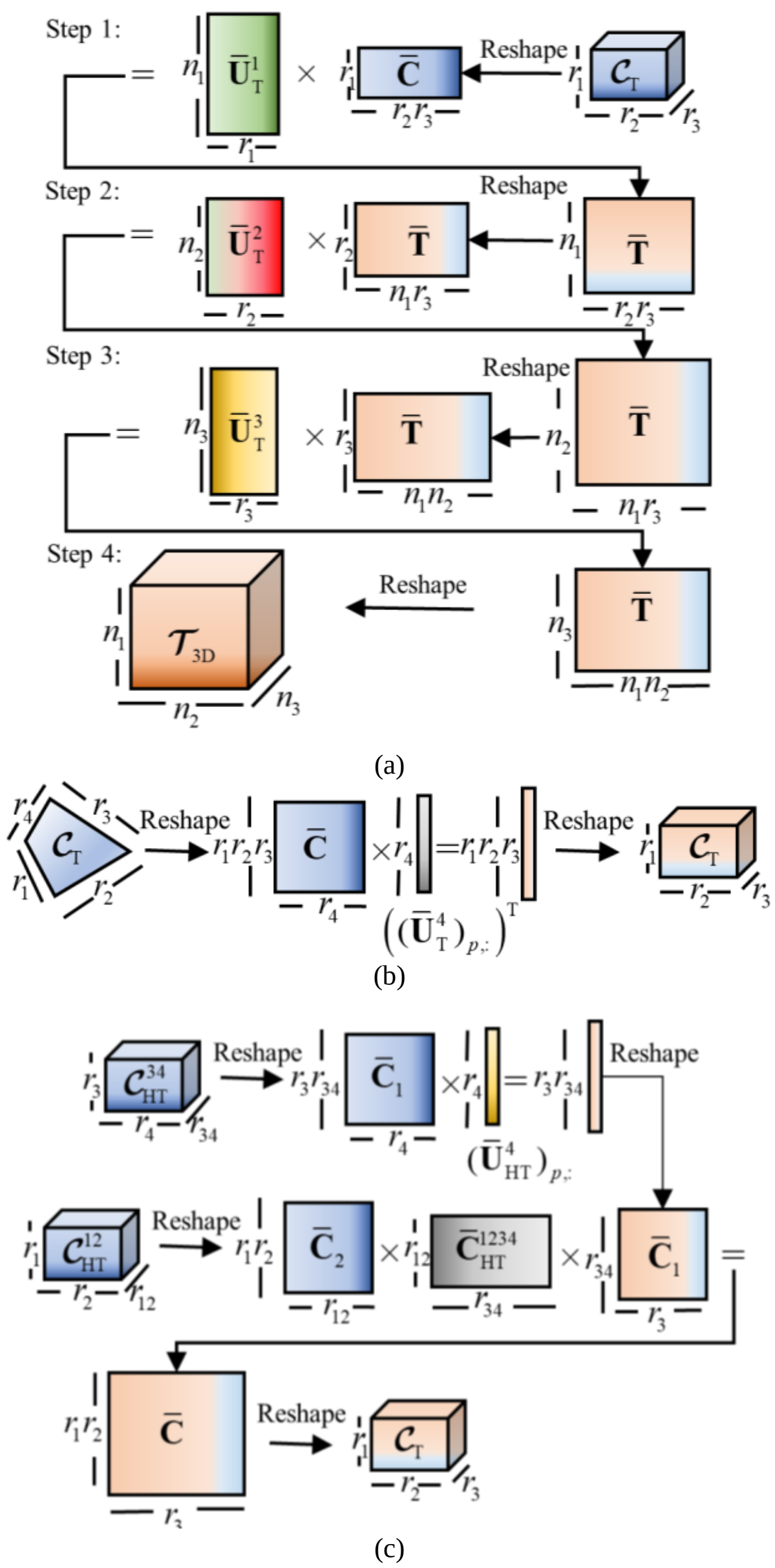

Fig. 3. (a) Tucker-3D restoration and additional steps for (b) Tucker-4D and (c) H-Tucker restorations.

*2) TT Restoration*

As in Tucker restoration scheme, the original FFT'ed translation operator tensor pertinent to $p^{\text{th}}$ plane-wave direction, $\mathcal{T}_{\mathbf{v}'-\mathbf{v}}(\hat{\mathbf{k}}_p) = \mathcal{T}_{3\text{D}}$, can be obtained from TT-compressed tensors via matrix product and reshaping operations, which are outlined in Figs. 4(a) and (b) for TT-3D and TT-4D methodologies, respectively. For TT-3D restoration [Fig. 4(a)], the core tensor $\mathcal{C}_{\text{TT}}^{1}$ is converted to a matrix $\bar{\mathbf{C}}$, which is multiplied by the factor matrix $\bar{\mathbf{U}}_{\text{TT}}^{1}$ in Step 1. The resulting auxiliary matrix $\bar{\mathbf{T}}_1$ is reshaped and then multiplied by the factor matrix $\bar{\mathbf{U}}_{\text{TT}}^{3}$ in Step 2. The resulting auxiliary matrix $\bar{\mathbf{T}}_1$ is converted to the tensor $\mathcal{T}_{3\text{D}}$ in Step 3.

For TT-4D restoration [Fig. 4(b)], the core tensor $\mathcal{C}_{\text{TT}}^{2}$ is converted to a matrix $\bar{\mathbf{C}}$, which is multiplied by the transpose of the factor matrix $\bar{\mathbf{U}}_{\text{TT}}^{3}$ in Step 1. In Step 2, $p^{\text{th}}$ column of the resulting auxiliary matrix $\bar{\mathbf{T}}_2$ is selected, reshaped, and then multiplied by the auxiliary matrix $\bar{\mathbf{T}}_1$, obtained at the end of Step 1 of the procedure in Fig. 4(a). The resulting auxiliary matrix $\bar{\mathbf{T}}_2$ is the converted to the tensor $\mathcal{T}_{\text{3D}}$ in Step 3.

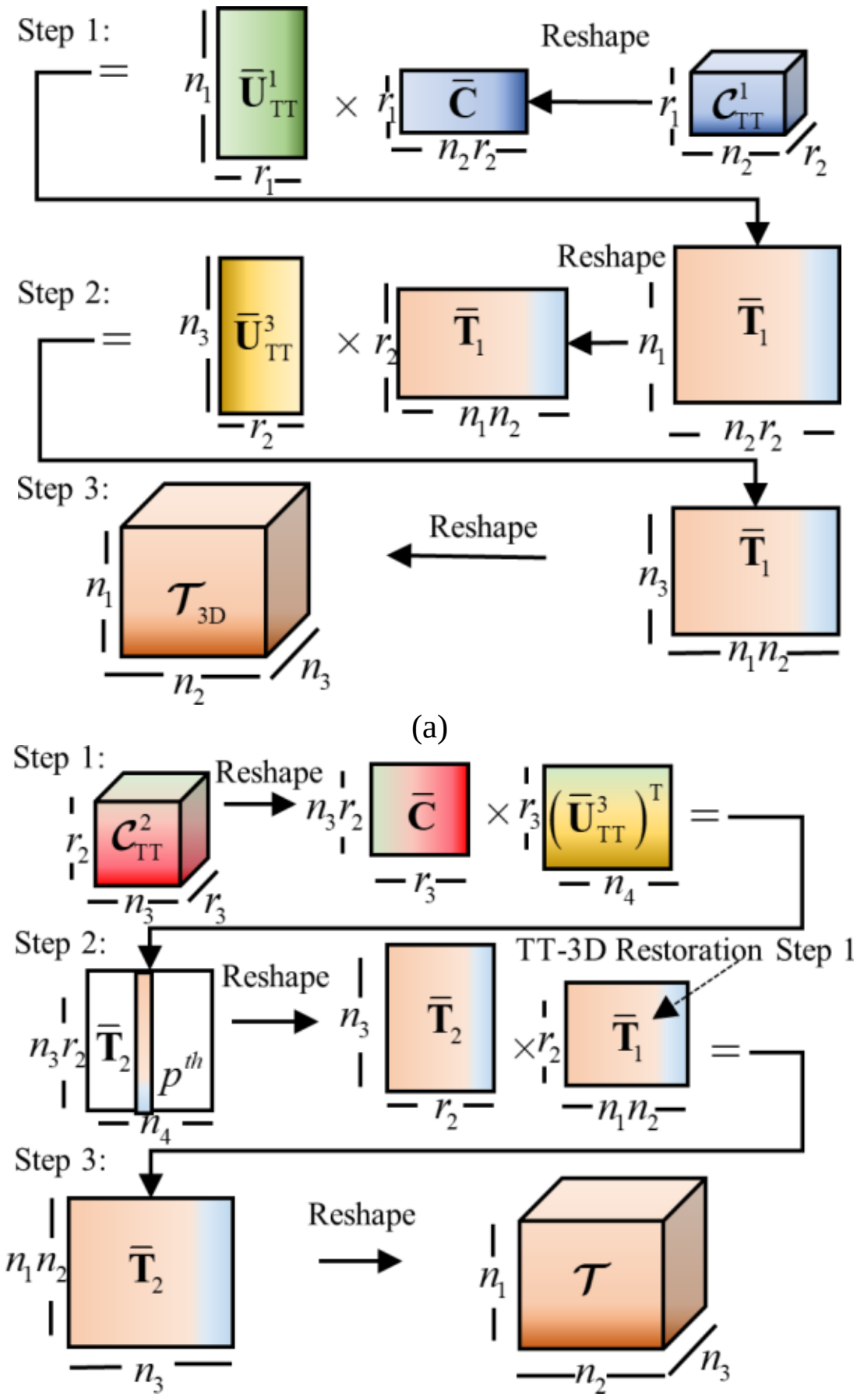

Fig. 4. (a) TT-3D and (b) TT-4D restoration.

## III. Numerical Results

In this section, the performances of the proposed tensor decomposition methodologies are extensively examined. In particular, their performances are first tested while compressing and decompressing/restoring the FFT'ed translation operator tensors generated for the EM scattering analysis of a sphere. In this test, the memory saving and computational overhead introduced by the proposed methodologies are quantified while the structure size, decomposition tolerance, FMM box size, FMM accuracy, and loss of the medium are changed. Next, the FMM-FFT-SIE simulator enhanced by the proposed tensor decompositions is used to analyze the EM scattering from a sphere and a plate. In these analyses, the accuracy, computational overhead, and memory saving introduced by the proposed methodologies are demonstrated.

In the tests below, the structures reside in free-space ($\varepsilon = \varepsilon_0$ and $\mu = \mu_0$), the frequency of analysis is 300 MHz, the FMM box size and FMM accuracy are set to $0.5\lambda$ and 5 digits, respectively, the tolerance for the decompositions $\gamma$ is $10^{-6}$, unless stated otherwise. The memory saving is computed by the ratio of the memory requirement of compressed tensors (i.e., core tensors and factor matrices) to that of original tensors. The computational overhead is defined as the ratio of the decompression time to the convolution time; while the decompression time is the CPU time required to restore the original tensors from the compressed tensors, the convolution time is the CPU time spent to perform convolutions (i.e., FFT, tensor-tensor multiply, and inverse FFT operations as in (5)) with the original tensors. All simulations are carried out on an Intel Xeon Gold 6142 CPU with 384 GB RAM.

### *A. The Performance of the Tensor Decompositions*

In the first example, the FFT'ed translation operator tensors generated during the EM scattering analysis of a sphere is considered. Unless stated otherwise, the sphere centered at the origin has the diameter of $64\lambda$.

*1) Structure Size:* Initially, the tensor decomposition methodologies are used to compress FFT'ed translation operator tensors while the sphere diameter is varied from $8\lambda$ to $64\lambda$. For such cases, the tensors computed for 435 plane-wave directions have dimensions $n_1 \times n_2 \times n_3 = 2K_x \times 2K_y \times 2K_z$ ranging from $32\times32\times32$ ($8\lambda$ case) to $256\times256\times256$ ($64\lambda$ case). The tensors are compressed via the proposed methodologies; the memory requirements of the compressed tensors and the original tensors are plotted with respect to total number of boxes $K = K_x K_y K_z$ [Fig. 5(a)]. Clearly, the memory requirements of the compressed tensors scale with $O(K^{\alpha})$, $0.55 \le \alpha \le 0.79$, while the memory requirement of the original tensors scales with $O(K)$. In Fig. 5(b), the CPU time to restore the original tensors from the compressed tensors for all plane-wave directions (i.e., decompression time) is plotted with respect to $K$. Furthermore, the CPU time required to perform convolutions with original tensors for each case is also included in Fig. 5(b). Clearly, the decompression times required by Tucker-3D and TT-3D methodologies are a fraction of the CPU time required by convolutions and scale with $O(K \log K)$. On the other hand, the decompression times required by H-Tucker and TT-4D are more than the CPU time required by convolutions and scale with $O(K^{0.79})$ and $O(K^{1.23})$, respectively. Although the CPU time scaling of the H-Tucker appears to be the best, its crossover with the CPU time scaling of the convolutions occurs at very large $K$ values. To this end, the Tucker-3D is favorable for small and large $K$ values, especially when the least decompression time is sought for. The memory saving and computational overhead introduced by each methodology are plotted [Figs. 5 (c) and (d)]. With increasing structure size (or $K$), the memory saving increases [Fig. 5(c)] and the computational overhead decreases. H-Tucker yields the largest memory saving and acceptable computational overhead.

In Table I, the memory saving and computational overhead introduced by each methodology are tabulated for $64\lambda$ case. Apparently, while H-Tucker yields the maximum memory saving with a moderate computational overhead, the Tucker-3D requires the minimum computational overhead. TT-4D yields the second-best memory saving with a decompression time more than the convolution time. On the other hand, TT-3D yields the worst memory saving with the computational overhead comparable to that of Tucker-3D.

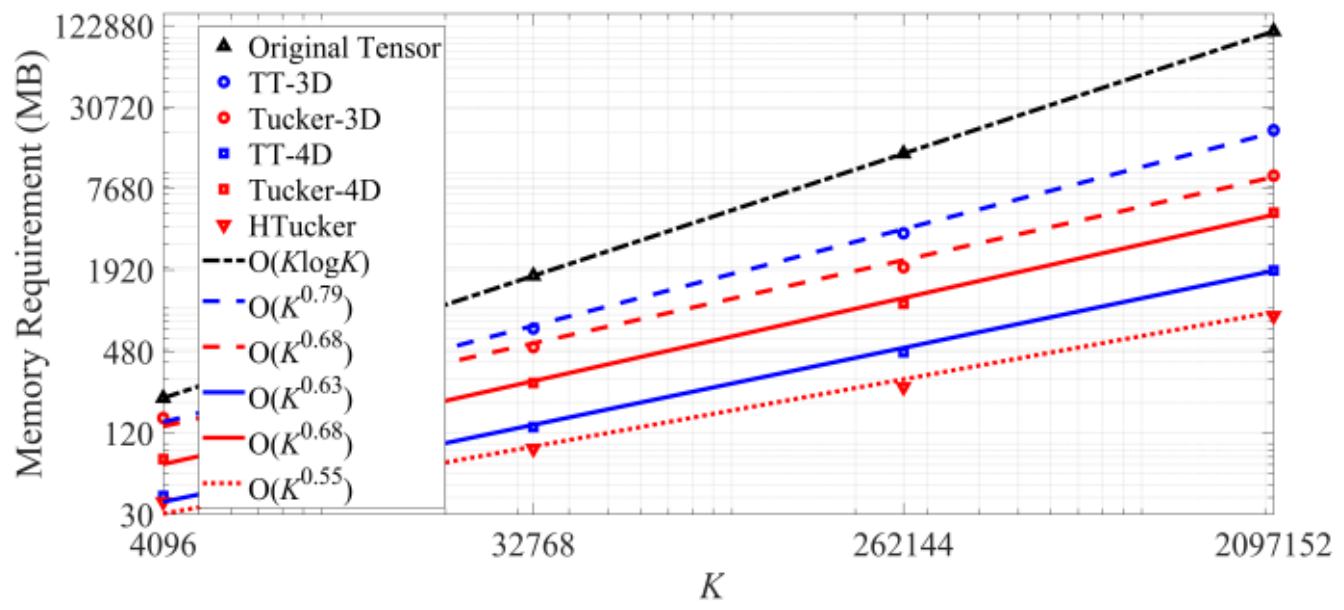

(a)

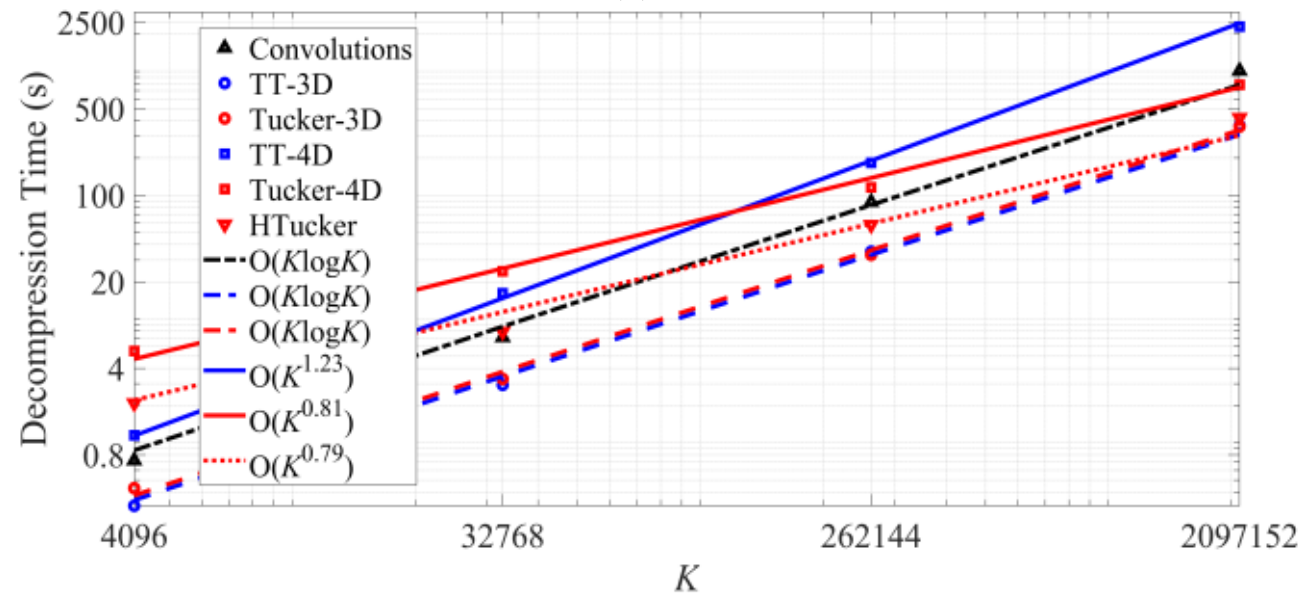

(b)

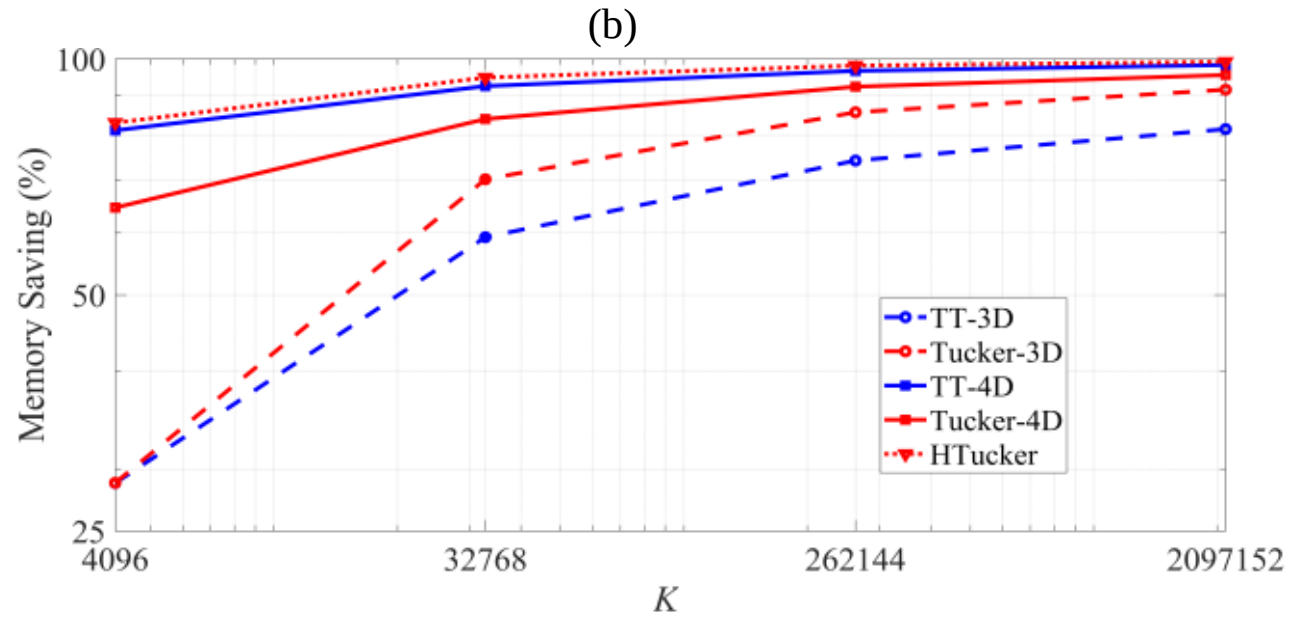

(c)

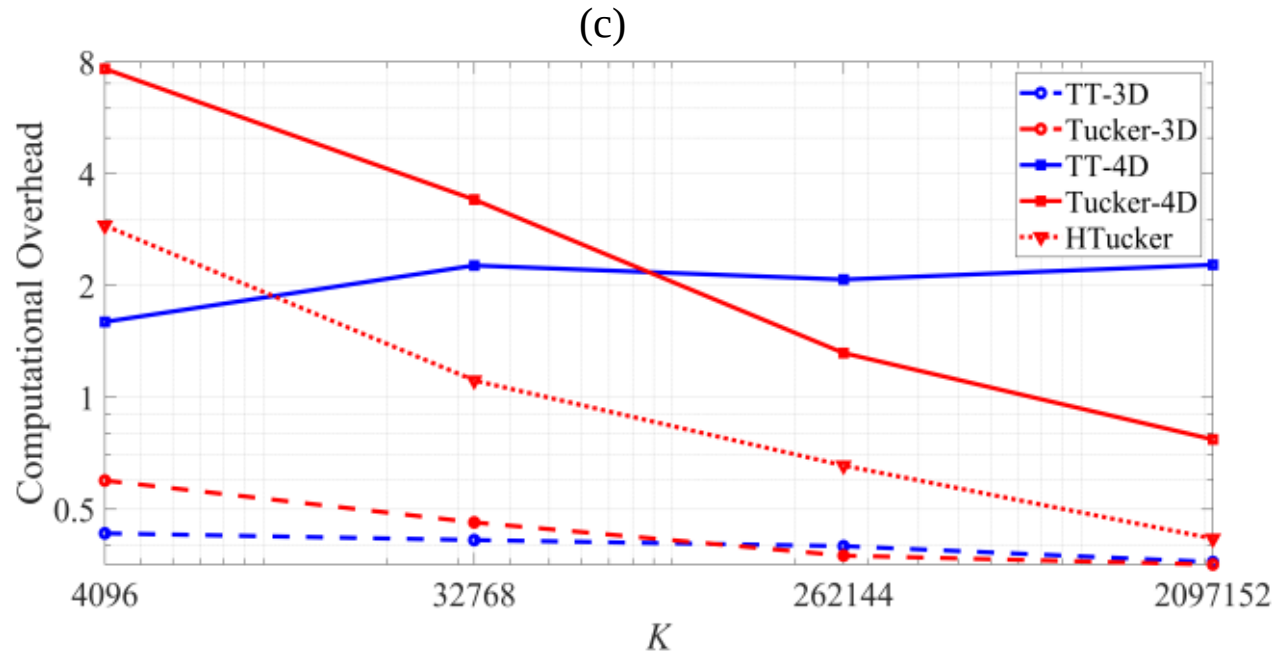

(d)

Fig. 5. (a) Memory requirement of the compressed and original tensors and (b) the CPU time for decompressing the FFT'ed translation operator tensors, (c) the memory saving, (d) the computational overhead via each methodology while the structure size is increased from $8\lambda$ ($K = 4096$) to $64\lambda$ ($K = 2,097,152$).

TABLE I

MEMORY SAVING AND COMPUTATIONAL OVERHEAD IN STRUCTURE SIZE TEST OF THE PEC SPHERE (64 $\lambda$ CASE)

| Methodology | Memory Saving (%) | Computational Overhead |
|---|---|---|
| TT-3D | 81.41 | 0.36 |
| Tucker-3D | 91.41 | 0.35 |
| TT-4D | 98.29 | 2.27 |
| Tucker-4D | 95.43 | 0.77 |
| H-Tucker | 99.21 | 0.42 |

The multilinear ranks of the compressed tensors with respect to increasing structure size are tabulated in Table II. For TT-3D and Tucker-3D, the maximum ranks of the compressed tensors generated for 435 plane-wave directions are listed. Clearly, the ranks are nearly the same for TT-3D and Tucker-3D methodologies and increase linearly with increasing structure. Furthermore, the ranks of the Tucker-4D/H-Tucker related to spatial dimensions (i.e., $r_1, r_2$, and $r_3$) are nearly the same as those of Tucker-3D while the ranks of the transfer matrix in H-Tucker (i.e., $r_{12}, r_{34}$,) are similar to $r_2$ in TT-4D. The ranks of Tucker-4D/H-Tucker and TT-4D pertinent to the plane-wave dimension, $r_4$ and $r_3$, are almost half of the number of plane-wave and do not change with increasing structure size.

TABLE II

RANKS OF COMPRESSED TENSORS FOR EACH METHODOLOGY WITH INCREASING STRUCTURE SIZE

| Methodology \ $n = n_1 \times n_2 \times n_3$ | | 32 | 64 | 128 | 256 |
|---|---|---|---|---|---|
| TT-3D | $r_{max}$ | 28 | 42 | 67 | 114 |
| Tucker-3D | $r_{max}$ | 29 | 44 | 69 | 115 |
| TT-4D ($r_3 = 225$) | $r_1$ | 27 | 42 | 67 | 113 |
| | $r_2$ | 322 | 503 | 826 | 1438 |
| Tucker-4D ($r_4 = 225$) | $r_1 = r_2 = r_3$ | 28 | 44 | 68 | 114 |
| H-Tucker ($r_4 = 225$) | $r_1 = r_2 = r_3$ | 25 | 43 | 63 | 113 |
| | $r_{12} = r_{34}$ | 327 | 511 | 836 | 1452 |

*2) Decomposition Tolerance:* Next, while the decomposition tolerance $\gamma$ is changed from $10^{-3}$ to $10^{-6}$, the memory saving and computational overhead introduced by each methodology are plotted [Figs. 6 (a) and (b)]. With decreasing decomposition tolerance (or relative error in the compressed representations), the memory saving decreases [Fig. 6(a)]. Clearly, 4D methodologies yield higher memory savings compared to the 3D methodologies. H-Tucker yields the maximum memory saving. Furthermore, the computational overhead increases with decreasing decomposition tolerance.

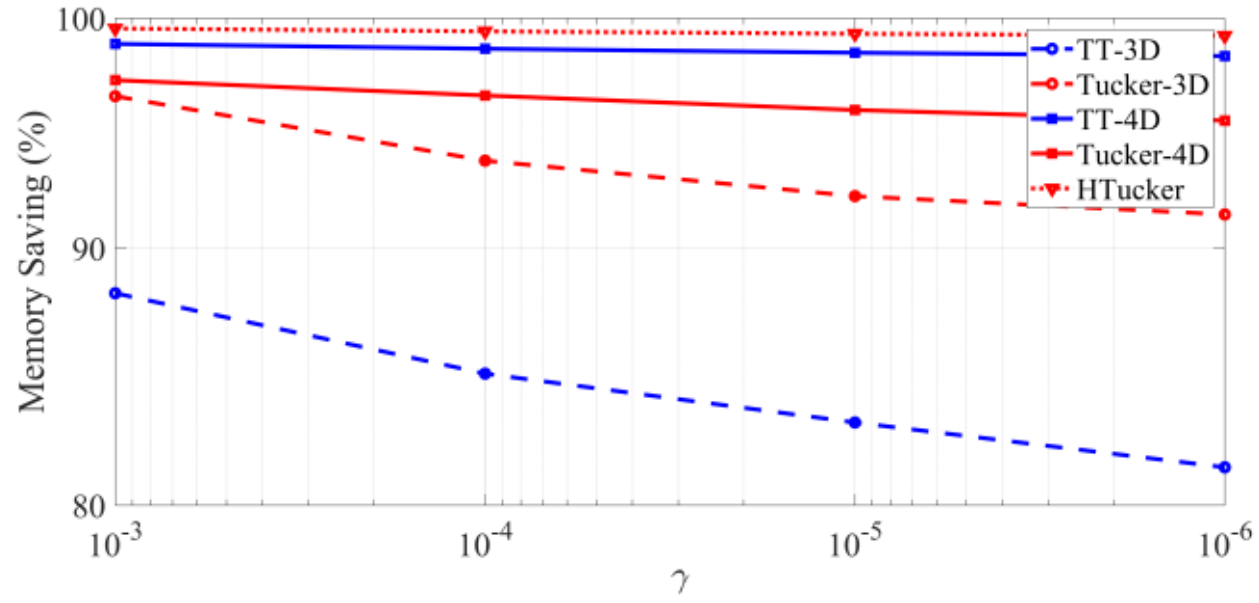

(a)

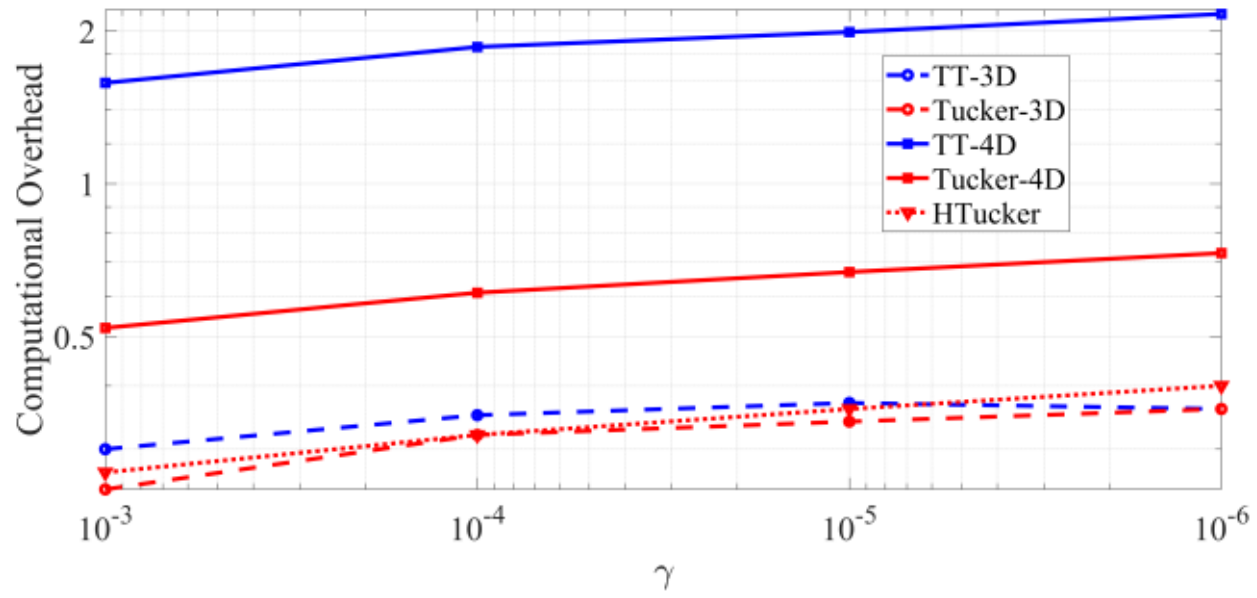

(b)

Fig. 6. (a) The memory saving and (b) the computational overhead w.r.t. $\gamma$ .

*3) FMM Box Size:* The performance of the proposed decomposition methodologies is examined when the FMM box size is set to $0.25\lambda$ , $0.33\lambda$ , $0.5\lambda$ , $0.66\lambda$ , and $\lambda$ for $32\lambda$ -diameter sphere. In this test, the dimensions of FFT'ed translation operator tensors change from $256\times256\times256$ for 231 plane-wave directions ( $0.25\lambda$ case) to $64\times64\times64$ for 1035 plane-wave directions ( $\lambda$ case). For each case, memory saving and computational overhead introduced by each methodology are plotted [Figs. 7(a) and (b)]. With decreasing box size, the memory saving increases and the computational overhead decreases. The 4D methodologies maintain high memory saving with the increase in the box size while the 3D methodologies cannot. For $0.25\lambda$ case, the memory saving and computational overhead associated with each methodology are tabulated [Table III]. All methodologies require decompression times less than convolution time. Furthermore, H-Tucker not only yields the maximum memory saving but also necessitates the minimal computational overhead. It shows H-Tucker has the best performance, especially when the compressibility of the translation operator tensors is high.

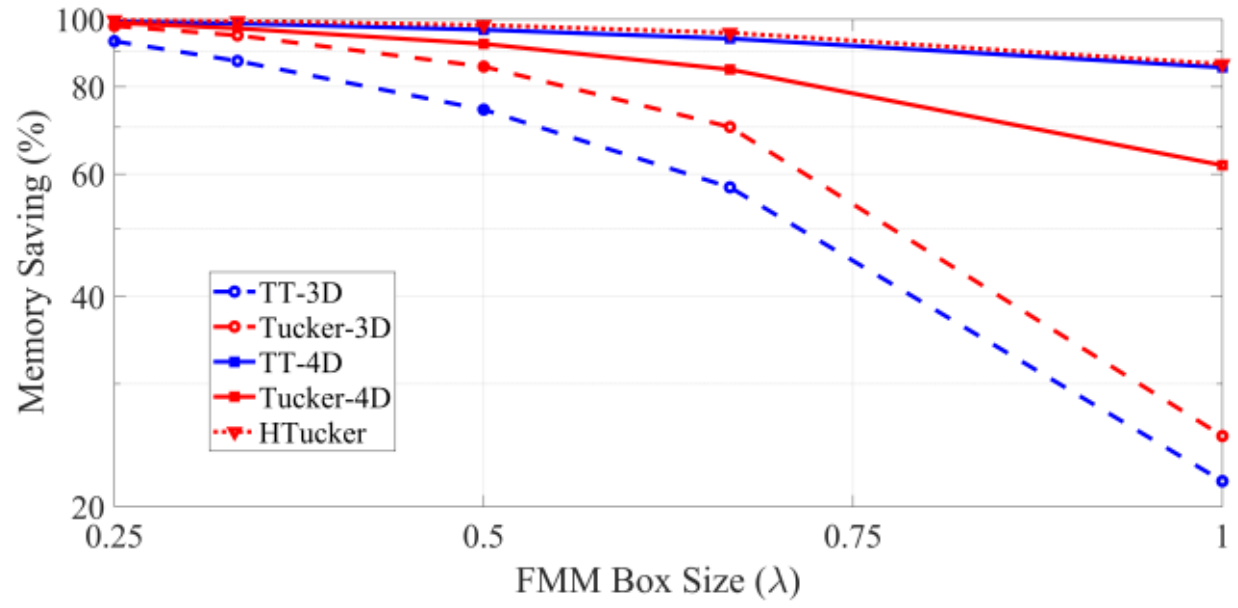

(a)

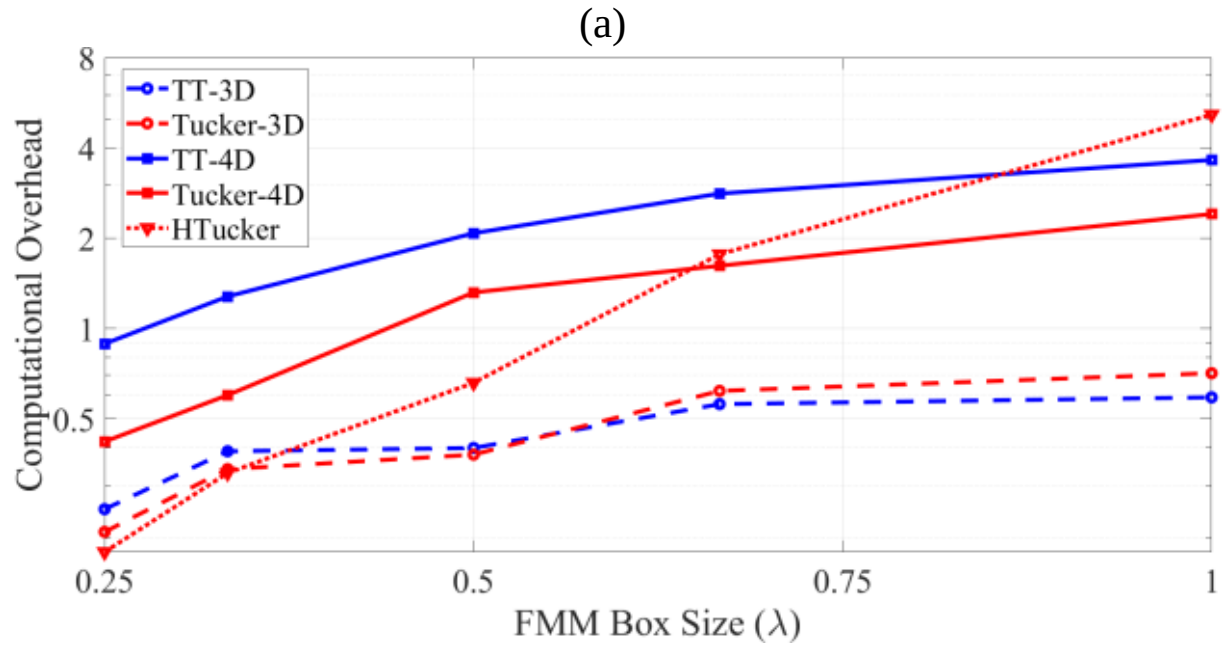

(b)

Fig. 7. (a) The memory saving and (b) the computational overhead w.r.t. FMM box size.

TABLE III

MEMORY SAVING AND COMPUTATIONAL OVERHEAD IN FMM BOX SIZE TEST OF THE PEC SPHERE ( $0.25\lambda$ BOX SIZE CASE)

| Methodology | Memory Saving (%) | Computational Overhead |
|---|---|---|
| TT-3D | 93.02 | 0.25 |
| Tucker-3D | 97.80 | 0.21 |
| TT-4D | 99.17 | 0.89 |
| Tucker-4D | 98.84 | 0.42 |
| H-Tucker | 99.75 | 0.18 |

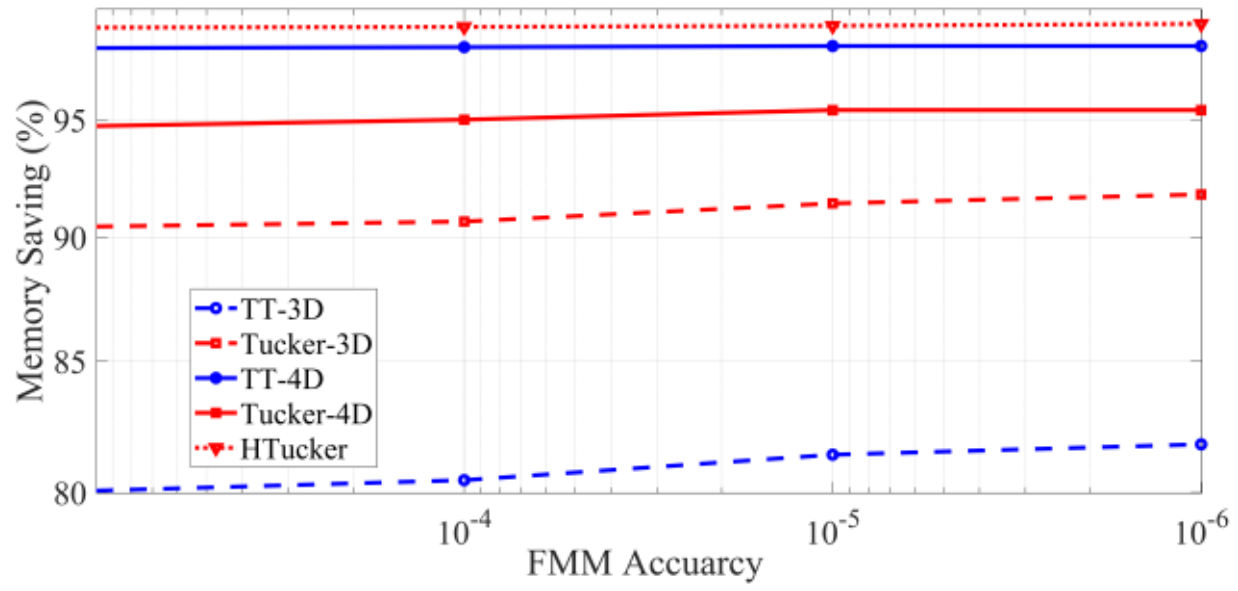

(a)

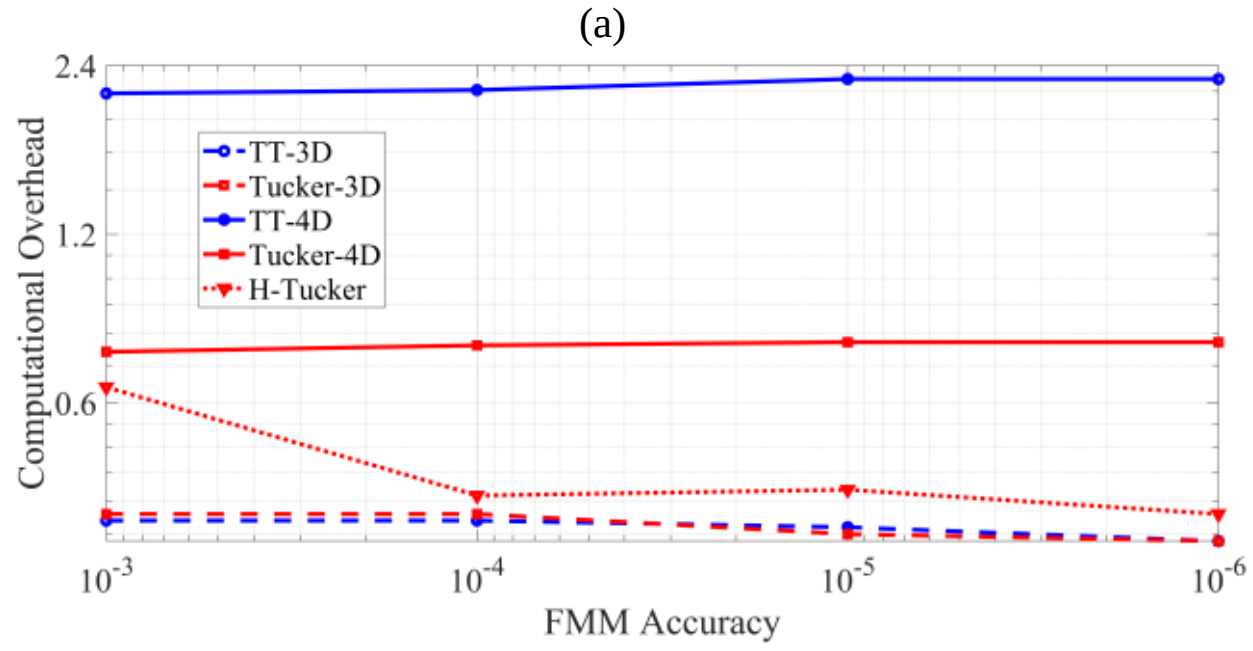

(b)

Fig. 8. (a) The memory saving and (b) the computational overhead w.r.t. FMM accuracy.

*4) FMM Accuracy:* Next, the performance of methodologies is investigated while the FMM accuracy varies from $10^{-3}$ to $10^{-6}$ ; this variation consecutively changes the number of plane-wave directions from 325 to 496. The memory saving and computational overhead associated with each methodology are plotted in Figs. 8(a) and (b), respectively. With increasing FMM accuracy, the memory saving slightly increases for 3D methodologies and is nearly constant for 4D methodologies. With increasing FMM accuracy, the computational overhead slightly decreases for 3D methodologies and increases for 4D methodologies. The increase in the accuracy of translation operator samples (i.e., plane-wave expansion of Green's function) results in a reduction in the rank. It is known that the increased accuracy in function samples on a grid reduces the rank in tensor decompositions. As a result, a decrease in the computational overhead (and an increase in the memory saving) for Tucker-3D and TT-3D can be observed. On the other hand, the slight increase in the computational overhead for 4D methodologies is due to the increase in the number of tensor entries along the plane-wave dimension, which increases with increasing FMM accuracy.

TABLE IV
THE PARAMETERS AND THE MAXIMUM MEMORY SAVING IN THE LOSS OF THE MEDIUM TEST

| Conductivity $\sigma$ ( S/m ) | 0 | 0.0167 | 0.0334 | 0.1335 |
|---|---|---|---|---|
| Relative Permittivity | 2 | $2-j$ | $2-2j$ | $2-8j$ |
| Number of Plane-Wave Directions | 703 | 703 | 861 | 1653 |
| Max. Memory Saving (%) | 96.74 | 98.22 | 99.98 | 99.99 |

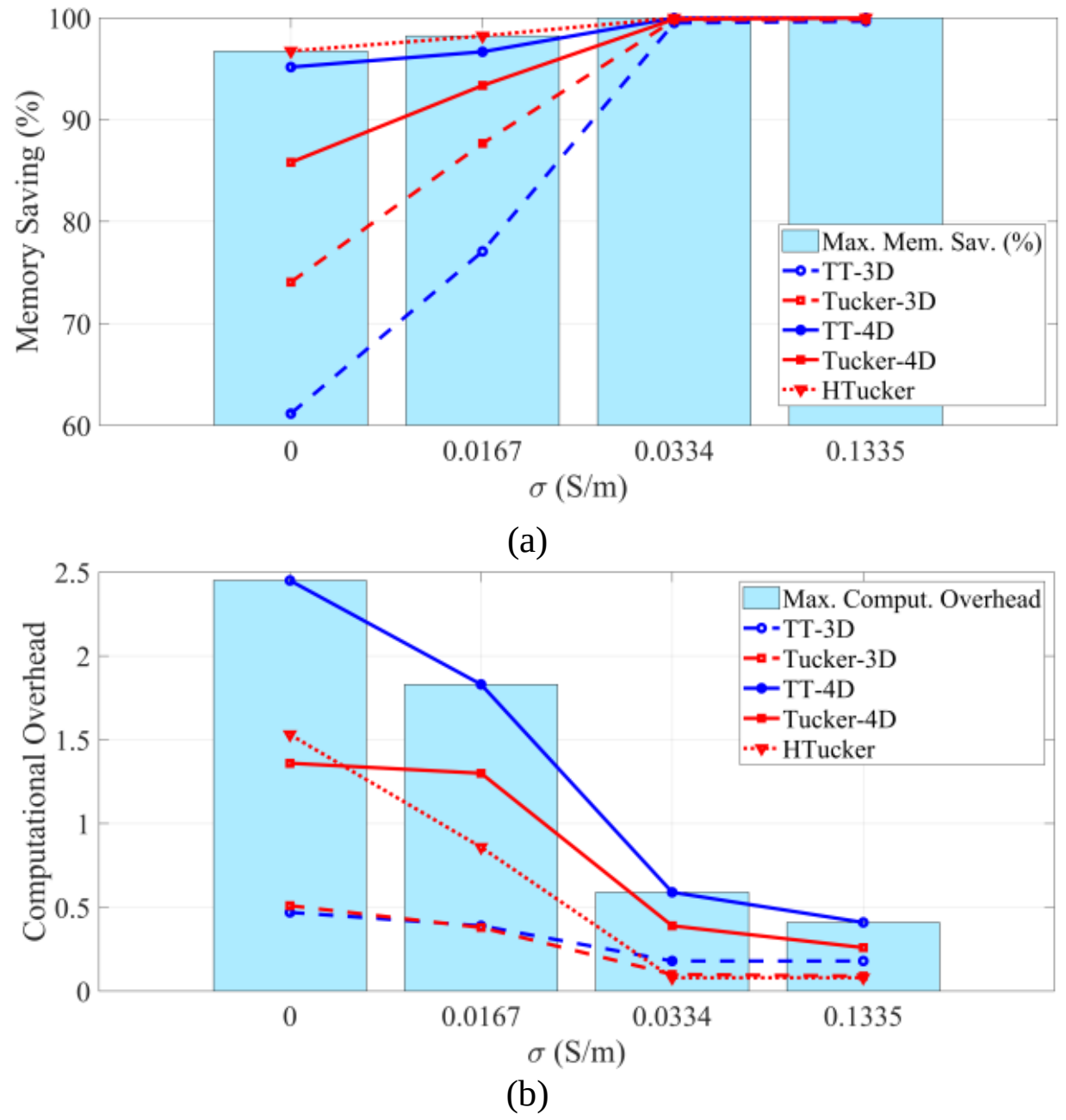

Fig. 9. (a) The memory saving and (b) computational overhead w.r.t. $\sigma$ .

5) *Loss of the Medium:* Last but not least, the performance of the proposed decompositions is quantified while increasing the loss of the medium housing a $32\lambda$ -diameter PEC sphere. The loss in the medium is introduced by changing the conductivity $\sigma$ from 0 to 0.1335 (or the complex relative permittivity from 2 to $2-8j$ ) [Table IV]. All methodologies perform better for the analysis with lossy medium compared to the analysis with lossless medium. This is not surprising as the matrix/tensor decomposition ranks reduce with decreasing EM interactions between basis functions in lossy media [34, 35]. The maximum memory saving tabulated in Table IV is again achieved by H-Tucker. The memory saving and computational overhead introduced by each methodology are plotted with respect to $\sigma$ in Fig. 9(a) and Fig. 9(b), respectively. With the increasing loss, the memory saving achieved by each decomposition increases while the computational overhead decreases and becomes a fraction of the convolution time. Achieved high memory saving and negligible computational overhead make the proposed tensor methodologies more appealing for the EM analysis of realistic scenarios involving the lossy media.

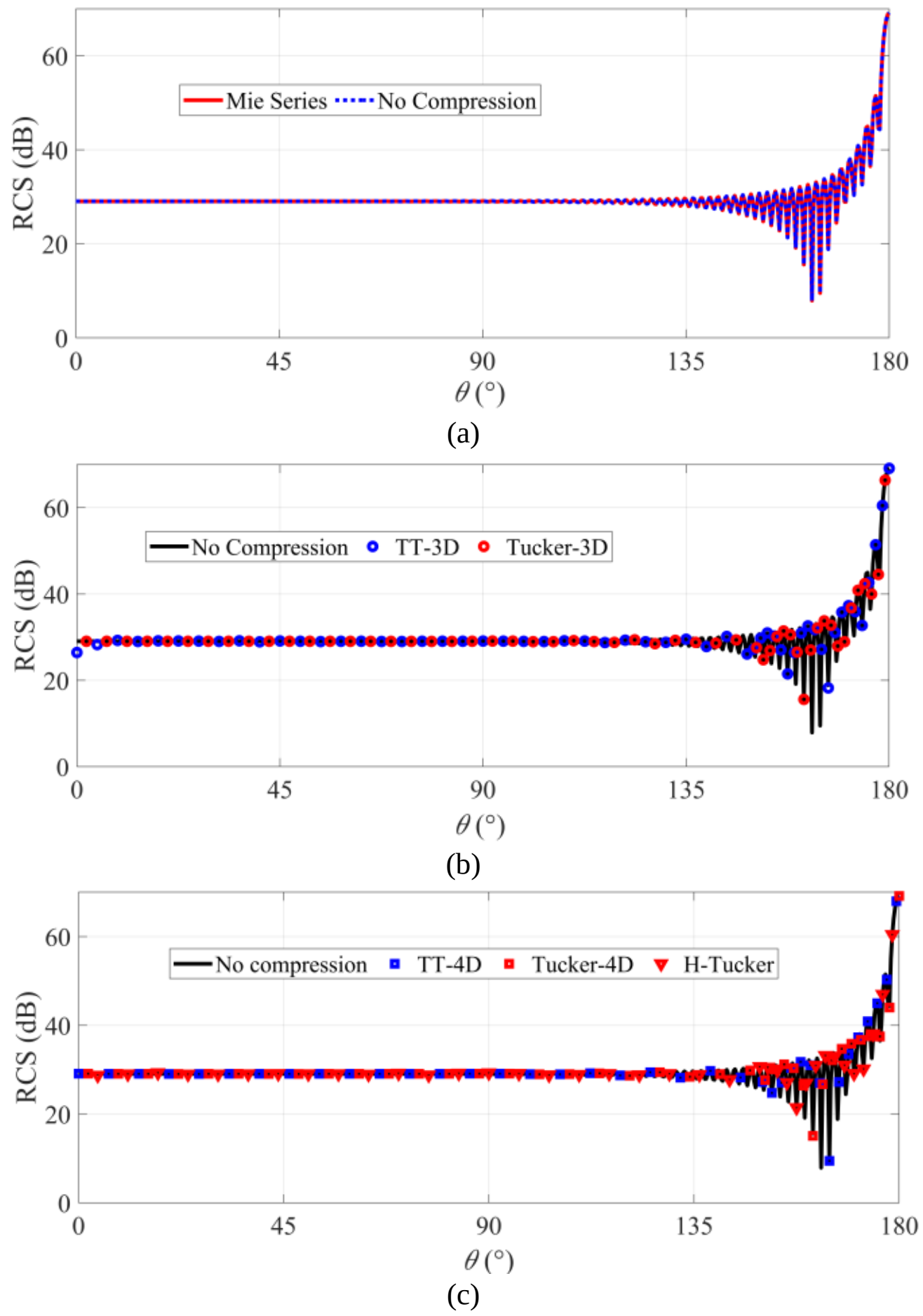

Fig. 10. The RCS of the PEC sphere obtained by (a) FMM-FFT-SIE simulator with no compression and Mie series, the FMM-FFT-SIE simulator enhanced with (b) 3D and (c) 4D methodologies and with no compression.

## B. *The Analysis of EM Scattering from a Sphere*

The proposed methodologies are used in an FMM-FFT-SIE simulator for the analysis of EM scattering from a $32\lambda$ -diameter PEC sphere. In this analysis, the sphere is discretized by 1,075,662 RWG basis functions and illuminated by a $y$ -polarized plane-wave propagating along $-z$ -direction. The bistatic radar cross section (RCS) of the sphere is computed along $\theta$ -direction while $\theta$ changes from $0°$ to $180°$ ( $\phi = 90°$ ). First, the RCS of the sphere computed by the FMM-FFT-SIE simulator without employing any compression scheme is compared with the RCS obtained by the Mie series solution [Fig. 10(a)]; a perfect agreement between results is observed. This validates the accuracy of the simulator. Next, the RCSs obtained by the Tucker-3D and TT-3D enhanced FMM-FFT-SIE simulator are compared with the original RCS computed by FMM-FFT-SIE simulator with no compression [Fig. 10(b)]. Likewise, the RCSs obtained by Tucker-4D, TT-4D, and H-Tucker enhanced FMM-FFT-SIE simulator are compared with the original RCS [Fig. 10(c)]. It is clear in Figs. 10 (b) and (c) that the Tucker-3D, Tucker-4D, TT-4D and H-Tucker enhanced simulator yields highly accurate results while the TT-3D enhanced simulator provides less accurate results. This can also be seen from the $L^2$ norm of the relative differences between the RCSs obtained by the tensor decomposition enhanced simulator and the original RCS in Table V. Furthermore, Table V also tabulates the memory saving and computational overhead associated with each methodology. Again, H-Tucker yields the maximum memory saving with an acceptable computational overhead while the Tucker-3D provides reasonable memory saving with the least computational overhead. TT-4D yields second-highest memory saving with a maximum computational overhead. TT-3D performs worst among all methodologies; it provides the minimum memory saving and the computational overhead, almost the same as that of Tucker-3D. For this example, three large data structures in FMM-FFT-SIE simulator, namely near-field interaction matrices, the matrices holding the far-field signatures, and FFT'ed translation operator tensors, require 14.4 GB, 20.7 GB, and 23.7 GB, respectively. Clearly, translation operators occupy the largest memory among these data structures. The total memory requirement of the simulator with temporary data structures is 63.1 GB. By using the H-Tucker decomposition, the total memory requirement of the simulator is reduced to 40GB (by a factor of 1.58) while introducing a computational penalty of 66%. It should be noted here that the peak memory requirement of the tensor decomposition enhanced FMM-FFT-SIE simulator is always less than that of the original FMM-FFT-SIE simulator. During the translation stage of each matrix-vector multiplication, the peak memory requirements of the Tucker-3D, TT-3D, Tucker-4D, TT-4D, and H-Tucker methodologies are 3.4 GB, 5.6 GB, 2.2 GB, 2.2 GB, and 1.3 GB, respectively. Note that the peak memory requirement at this stage of the original FMM-FFT-SIE simulators is 24.5 GB. While obtaining the tensor decompositions in the setup stage, the peak memory requirements of the Tucker-3D and TT-3D are 0.6 GB and 0.5 GB, respectively. Those of Tucker-4D, TT-4D, and H-Tucker methodologies are 46.7, 47.6, and 58.4 GB, respectively, which are reduced to 24.7 GB using the cross-based algorithms [26-29].

TABLE V

THE ACCURACY, MEMORY SAVING, AND COMPUTATIONAL OVERHEAD INTRODUCED BY METHODOLOGIES FOR THE PEC SPHERE EXAMPLE

| Methodology | $L^2$ -Relative Difference Norm (%) | Memory Saving(%) | Computational Overhead |
|---|---|---|---|
| TT-3D | 0.81 | 74.22 | 0.40 |
| Tucker-3D | 0.01 | 85.54 | 0.38 |
| TT-4D | 0.04 | 96.60 | 2.08 |
| Tucker-4D | 0.02 | 92.22 | 1.32 |
| H-Tucker | 0.06 | 98.14 | 0.66 |

## *C. The Analysis of EM Scattering from a Plate*

Finally, the proposed decompositions are used in the FMM-FFT-SIE simulator for the EM scattering analysis of a PEC plate with dimensions $30\lambda \times 42\lambda$ . The plate is diagonally centered at the origin [Fig. 11(a)], discretized by 377,280 RWG basis functions, and excited by an $x$ -polarized plane-wave propagating along $-z$ -direction. The bistatic RCS of the PEC plate is computed along $\theta$ -direction while $\theta$ changes from $45°$ to $225°$ and $\phi = 90°$ . The RCS computed by the FMM-FFT-SIE simulator without compression is compared with the RCS obtained by the analytical formula [36] [Fig. 11(a)]. A good agreement between results is observed; the discrepancy around $\theta = 135°$ is expected as the analytical formula does not take into account the edge diffraction effects. Next, the RCSs obtained by the TT-3D, Tucker-3D, TT-4D, Tucker-4D, and H-Tucker decomposition enhanced FMM-FFT-SIE simulator are compared with the RCS obtained by the FMM-FFT-SIE simulator without compression [Figs. 11(b)-(c)]. An excellent agreement between results is observed. The accuracy of the RCS obtained by FMM-FFT-SIE simulator enhanced by each methodology as well as the memory saving and computational overhead associated with each methodology are tabulated in Table VI. It is demonstrated again that H-Tucker performs the best in memory saving comparison with a decompression time less than the convolution time. For this example, three large data structures in FMM-FFT-SIE simulator, namely near-field interaction matrices, the matrices of the far-field signatures, and the FFT'ed translation operator tensors, require 5.3 GB, 6.5 GB, and 16.7 GB memory, respectively. The total memory requirement of the simulator with the temporary data structure is 30 GB. By using the H-Tucker decomposition, the total memory requirement of the simulator is reduced to 14 GB (by a factor of 2.14) while introducing a computational penalty of 41%.

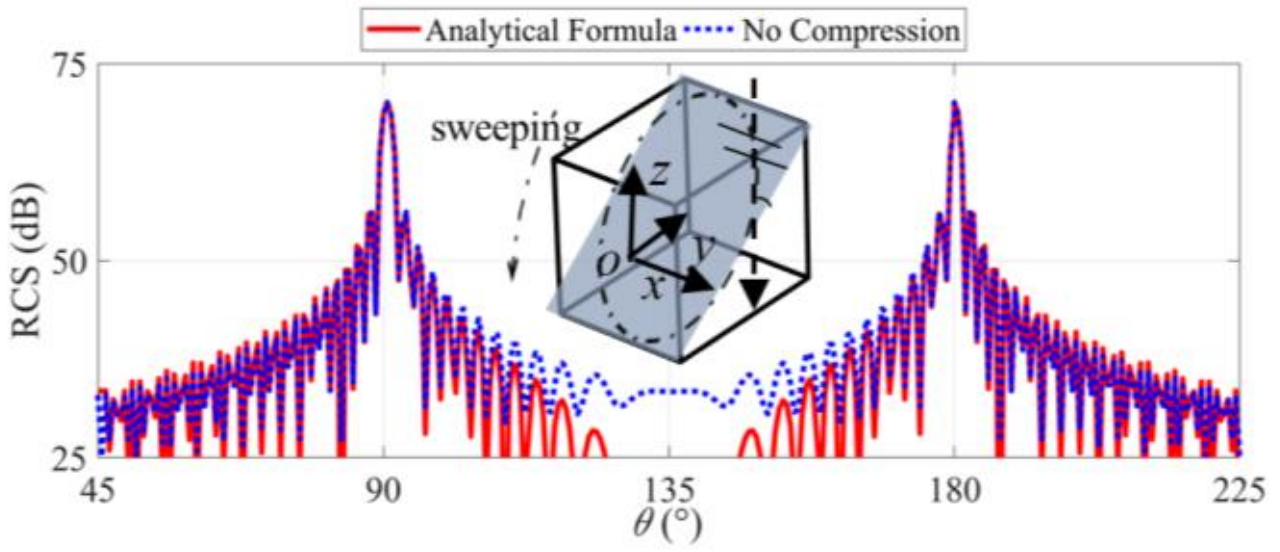

(a)

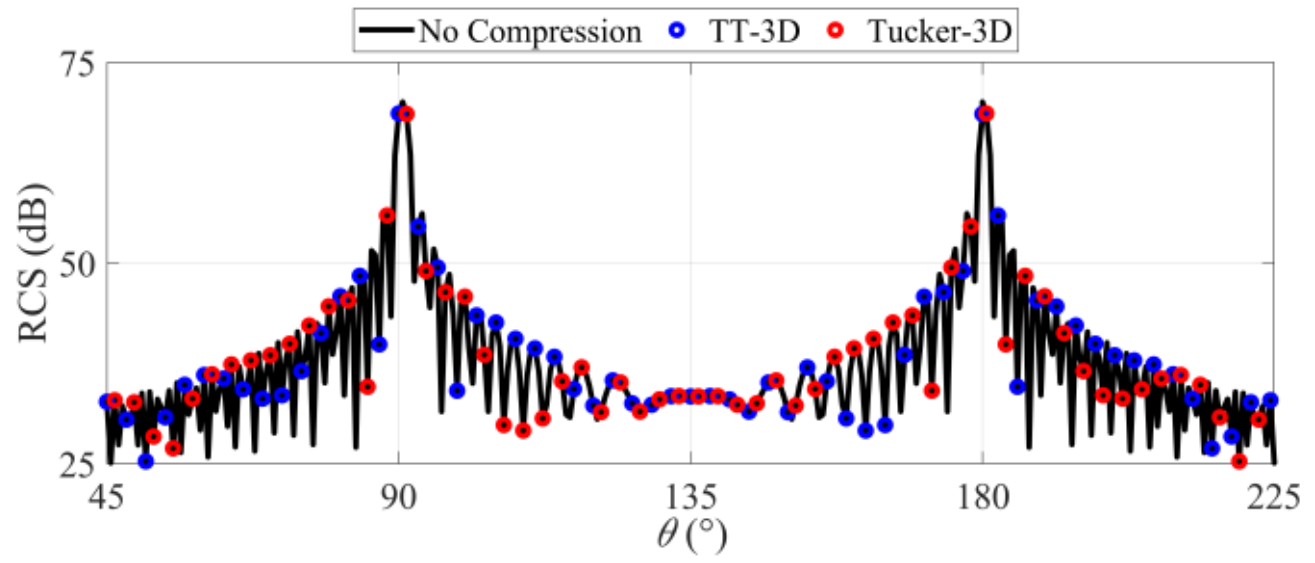

(b)

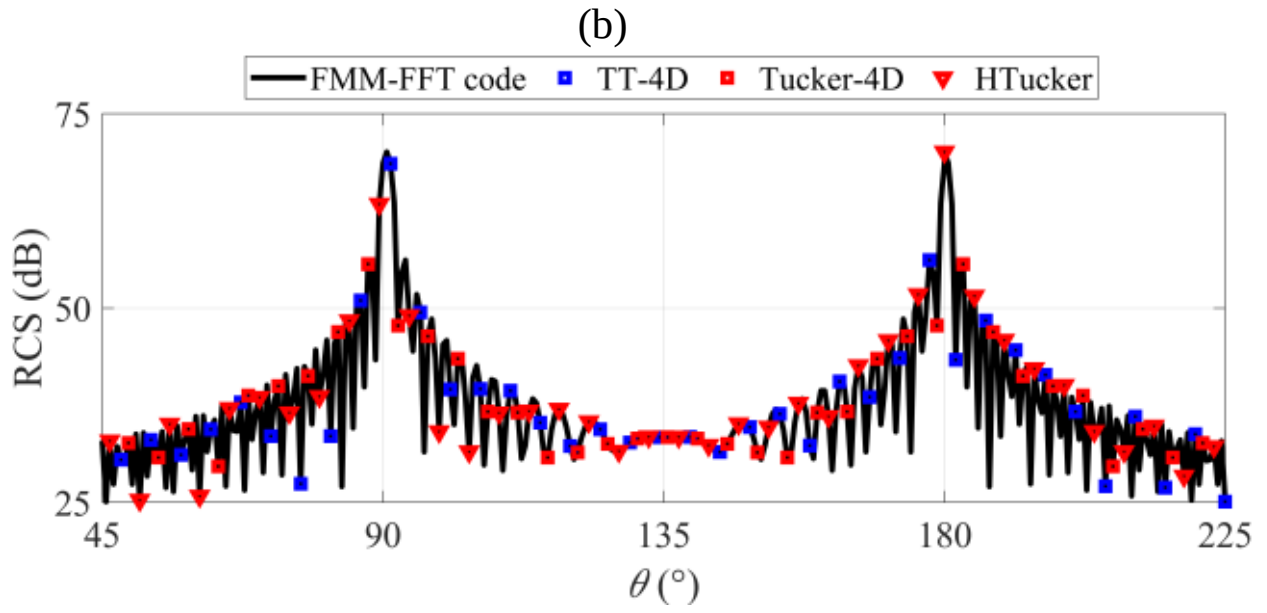

(c)

Fig. 11. The RCS of the PEC plate obtained by (a) FMM-FFT-SIE simulator with no compression and analytical formula, the FMM-FFT-SIE simulator enhanced with (b) 3D and (c) 4D methodologies and with no compression.

TABLE VI

THE ACCURACY, MEMORY SAVING, AND COMPUTATIONAL OVERHEAD INTRODUCED BY METHODOLOGIES FOR THE PEC PLATE EXAMPLE

| Methodology | $L^2$ -Relative Difference Norm (%) | Memory Saving (%) | Computational Overhead |
|---|---|---|---|
| TT-3D | 0.02 | 76.71 | 0.25 |
| Tucker-3D | 0.02 | 87.38 | 0.27 |
| TT-4D | 0.01 | 96.57 | 1.13 |
| Tucker-4D | 0.03 | 93.38 | 0.75 |
| H-Tucker | 0.03 | 98.18 | 0.41 |

## IV. CONCLUSION

In this paper, Tucker, H-Tucker, and TT based tensor decomposition methodologies were proposed for reducing the memory requirement of the FFT'ed translation operator tensors. These decompositions can be readily implemented in the existing codes of the FMM-FFT-SIE simulators and yield significant memory saving while imposing negligible/tolerable computational overhead. The proposed decomposition methodologies allow boosting the applicability of the FMM-FFT-SIE simulators on the limited computational resources. Detailed performance analysis of these methodologies for different simulation parameters such as structure size, decomposition tolerance, FMM box size, FMM accuracy, and medium loss was provided. In this analysis, H-Tucker yielded the maximum memory saving, while the Tucker-3D introduced the least computational overhead. While the TT-3D performed worst in all tests, Tucker-4D/H-Tucker yielded moderate/acceptable computational overhead but more memory saving compared to the 3D methodologies. Currently, the application of the proposed methodologies to the compression of far-fields in the FMM-accelerated EM simulators is investigated [37, 38].

## APPENDIX
## ALGORITHMS FOR OBTAINING TENSOR DECOMPOSITIONS

In this section, the classical SVD-based algorithms for obtaining Tucker (Algorithm 1), H-Tucker (Algorithm 2), and TT (Algorithm 3) decompositions are provided.

**Algorithm 1**: Tucker-SVD

1: **Inputs:** $d$ -dimensional array $\mathcal{T} = \{\mathcal{T}_{3\text{D}}, \mathcal{T}_{4\text{D}}\}$ and $\gamma$ .
2: **Outputs:** $\mathcal{C}_{\text{T}}$ and $\bar{\mathbf{U}}^i_{\text{T}}$, $i = 1,\ldots,d$ .
3: **Initialize:** $\mathcal{C}_{\text{T}} = \mathcal{T}$ .
4: **for** $i = 1:d$ **do**
5: obtain mode $-i$ unfolding matrix of $\mathcal{T}$ , $\bar{\mathbf{T}}^i$ .
6: compute SVD of $\bar{\mathbf{T}}^i$ as $\bar{\mathbf{T}}^i = \bar{\mathbf{U}}^i \bar{\mathbf{\Sigma}}^i \bar{\mathbf{V}}^{i*}$ .
7: assign the index of maximum (normalized) singular value in $\bar{\mathbf{\Sigma}}^i$ smaller than $\gamma / d^{0.5}$ as $r_i$ , truncate $\bar{\mathbf{U}}^i$ with $r_i$ , and assign $\bar{\mathbf{U}}^i_{\text{T}} = \bar{\bar{\mathbf{U}}}^i$ .
8: $\mathcal{C}_{\text{T}} = \mathcal{C}_{\text{T}} \times_i \bar{\mathbf{U}}^{i*}_{\text{T}}$ .
9: **end for**
10: **return**

**Algorithm 2**: H-Tucker-SVD

1: **Inputs:** $\mathcal{T}_{4\text{D}}$ and $\gamma$ .
2: **Outputs:** $\mathcal{C}^{12}_{\text{HT}}$ , $\mathcal{C}^{34}_{\text{HT}}$ , $\bar{\mathbf{C}}^{1234}_{\text{HT}}$ , and $\bar{\mathbf{U}}^i_{\text{HT}}$ , $i = 1,\ldots,4$ .
3: obtain $\bar{\mathbf{U}}^i_{\text{HT}}$ , $i = 1,\ldots,4$ , via Algorithm 1.
4: form vector via Kronecker product
$$\text{vec}\left(\bar{\mathbf{W}}^1\right) = \left(\bar{\mathbf{U}}^1_{\text{HT}} \otimes \bar{\mathbf{U}}^2_{\text{HT}} \otimes \bar{\mathbf{U}}^3_{\text{HT}} \otimes \bar{\mathbf{U}}^4_{\text{HT}}\right) \text{vec}\left(\mathcal{T}_{4\text{D}}\right).$$
5: obtain matrix $\bar{\mathbf{W}}^{12}$ via matricization of $\bar{\mathbf{W}}^1$ [25].
6: compute SVD of $\bar{\mathbf{W}}^{12}$ as $\bar{\mathbf{W}}^{12} = \bar{\mathbf{C}}^{12} \bar{\mathbf{\Sigma}}^{12} \bar{\mathbf{V}}^{12*}$ .
7: truncate $\bar{\mathbf{C}}^{12}$ with $r_{12}$ via tolerance $\gamma / \sqrt{6}$ and convert to $\mathcal{C}^{12}_{\text{HT}}$ .
8: repeat steps 5-8 for $\mathcal{C}^{34}_{\text{HT}}$
9: form $\text{vec}\left(\bar{\mathbf{W}}^0\right) = \left(\bar{\mathbf{C}}^{12}_{\text{HT}} \otimes \bar{\mathbf{C}}^{34}_{\text{HT}}\right) \text{vec}\left(\bar{\mathbf{W}}^1\right)$ .
10: convert $\bar{\mathbf{W}}^0$ to $\bar{\mathbf{C}}^{1234}_{\text{HT}}$ .
11: **return**

**Algorithm 3**: TT-SVD

1: **Inputs:** $d$ -dimensional array $\mathcal{T} = \{\mathcal{T}_{3\text{D}}, \mathcal{T}_{4\text{D}}\}$ and $\gamma$ .
2: **Outputs:** $\mathcal{C}^i_{\text{TT}}$ and $\bar{\mathbf{U}}^{1,3}_{\text{TT}}$ , $i = 1,\ldots,d-2$ .
3: **Initialize:** obtain unfolding matrix of $\mathcal{T}$ , $\bar{\mathbf{C}}$ , with dimensions $r_0 n_1 \times n_2 \ldots n_d$ , where $r_0 = 1$ .
4: **for** $i = 1:d-1$ **do**
5: convert the dimensions of $\bar{\mathbf{C}}$ from $r_{i-1} \times n_i \ldots n_d$ to $r_{i-1} n_i \times n_{i+1} \ldots n_d$ (for $i \geq 2$ ).
6: compute SVD of $\bar{\mathbf{C}}$ as $\bar{\mathbf{C}} = \bar{\mathbf{U}}^i \bar{\mathbf{\Sigma}}^i \bar{\mathbf{V}}^{i*}$ .
7: assign the index of maximum (normalized) singular value in $\bar{\mathbf{\Sigma}}^i$ smaller than $\gamma / (d-1)^{0.5}$ as $r_i$ , truncate $\bar{\mathbf{U}}^i$ with $r_i$ .
7: obtain new core: $\mathcal{C}^{i-1}_{\text{TT}}$ with dimensions $r_{i-1} \times n_i \times r_i$ from $\bar{\mathbf{U}}^i$ with dimensions $r_{i-1} n_i \times r_i$ .
8: $\bar{\mathbf{C}} = \bar{\mathbf{\Sigma}}^i \bar{\mathbf{V}}^{i*}$ .
9: **end for**
10: $\bar{\mathbf{U}}^1_{\text{TT}} = \mathcal{C}^0_{\text{TT}}$ , $\bar{\mathbf{U}}^3_{\text{TT}} = \bar{\mathbf{C}}$ .
11: **return**

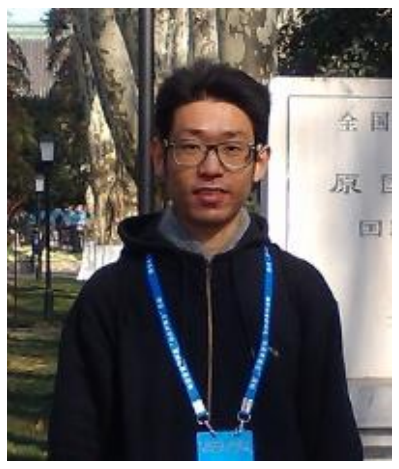

**Cheng Qian** received the B.S. and Ph.D. degree in electronics engineering from Nanjing University of Science and Technology, Jiangsu, China, in 2009 and 2015. From 2016 to 2018, he was a Research Associate with the Department of Applied Physics, The Hong Kong Polytechnic University, Hong Kong. Since 2019, he has been a Postdoctoral Researcher with the School of Electrical and Electronic Engineering, Nanyang Technological University, Singapore. His current research interests include computational electromagnetics and nonlinear plasmonics.

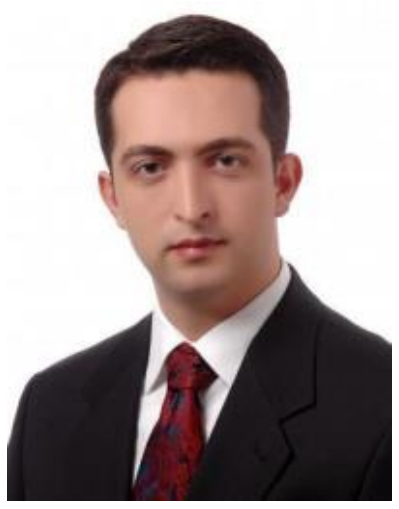

**Abdulkadir C. Yucel** (M’19-SM’20) received the B.S. degree in electronics engineering (*Summa Cum Laude*) from Gebze Institute of Technology, Kocaeli, Turkey, in 2005, and the M.S. and Ph.D. degrees in electrical engineering from the University of Michigan, Ann Arbor, MI, USA, in 2008 and 2013, respectively.

From September 2005 to August 2006, he worked as a Research and Teaching Assistant at Gebze Institute of Technology. From August 2006 to April 2013, he was a Graduate Student Research Assistant at the University of Michigan. Between May 2013 and December 2017, he worked as a Postdoctoral Research Fellow at the University of Michigan, Massachusetts Institute of Technology, and King Abdullah University of Science and Technology. Since 2018, he has been working as an Assistant Professor at the School of Electrical and Electronic Engineering, Nanyang Technological University, Singapore.

Dr. Yucel received the Fulbright Fellowship in 2006, Electrical Engineering and Computer Science Departmental Fellowship of the University of Michigan in 2007, and Student Paper Competition Honorable Mention Award at IEEE AP-S in 2009. He has been serving as an Associate Editor for the International Journal of Numerical Modelling: Electronic Networks, Devices and Fields and as a reviewer for various technical journals.